\documentclass[10pt]{article}
\usepackage{amsmath, latexsym, amssymb, amscd,amsbsy}

\numberwithin{equation}{section}
\newcommand{\eeq}{\end{equation}}
\newcommand{\eeqs}{\end{equation*}}
\newcommand{\bdm}{\begin{displaymath}}
\newcommand{\edm}{\end{displaymath}}
\newcommand{\beqs}{\begin{equation*}}
\newcommand{\beq}{\begin{equation}}

\newcommand{\bi}{\begin{itemize}}
\newcommand{\ei}{\end{itemize}}

\newcommand{\Tr}{\mathrm{Tr}}  
\newcommand{\tens}{\otimes}            

\newtheorem{theorem}{Theorem}[section]
\newtheorem{definition}[theorem]{Definition}
\newtheorem{lemma}[theorem]{Lemma}

\newtheorem{proposition}[theorem]{Proposition}
\newtheorem{corollary}[theorem]{Corollary}

\def\patrat{{\vcenter{\vbox{\hrule height.4pt \hbox{\vrule width.4pt 
height1.45ex \kern1.45ex \vrule width.4pt}
\hrule height.4pt}}}}

\def\qed{\hfill$\patrat$}

\newcommand{\barint}
{\hbox{$\int$\kern-0.75\intwidth\vrule width 0.5\intwidth height 2.4pt 
depth -2pt\kern0.25\intwidth}}
\newlength\intwidth
\setbox0=\hbox{$\int$}
\intwidth=\wd0

\newtheorem{ex}{Example}[section]
\def\iA{\mathcal A}
\def\iF{\mathcal F}
\def\iE{\mathcal E}
\newcommand{\<}{\langle}
\renewcommand{\>}{\rangle}

\begin{document}
\title{Local asymptotic normality in quantum statistics}
\author{M\u{a}d\u{a}lin Gu\c{t}\u{a}$^{1}$ and Anna Jen\v cov\' a$^{2}$\\ \\
$^{1}$ University of Nottingham, School of Mathematical Sciences\\
University Park, Nottingham NG7 2RD, U.K. \\ \\
$^{2}$ Mathematical Institute of the Slovak Academy of Sciences\\
 Stefanikova 49, 814 73 Bratislava, Slovakia
}
\date{}

\maketitle

\begin{center}
{\it
Dedicated to Slava Belavkin on the occasion of  his 60th anniversary}
\end{center}
\begin{abstract}
The theory of local asymptotic normality for quantum statistical experiments
is developed in the spirit of the classical result  from mathematical
statistics due to Le Cam.
 Roughly speaking, local asymptotic normality means that the family
$\varphi_{\theta_{0}+ u/\sqrt{n}}^{n}$ consisting of joint states of $n$  identically prepared quantum systems approaches in a statistical sense a family of Gaussian state $\phi_{u}$ of an algebra of canonical commutation relations. The convergence holds for all ``local parameters'' $u\in \mathbb{R}^{m}$ such that $\theta=\theta_{0}+ u/\sqrt{n}$ parametrizes a neighborhood of a fixed point $\theta_{0}\in \Theta\subset\mathbb{R}^{m}$.

In order to prove the result we define weak and strong convergence of quantum statistical experiments which extend  to the asymptotic framework the notion of quantum sufficiency introduces by Petz. Along the way we introduce the concept of canonical state of a statistical experiment, 
and investigate the relation between the two notions of convergence. For reader's convenience and completeness we review the relevant results of the classical as well as the quantum theory.
\end{abstract}

\section{Introduction}

The statistical interpretation of quantum mechanics, also known as the Born rule, is an interface connecting the mathematical framework based on Hilbert space operators and wave functions, with the reality in the form of measurement results. While the Born rule describes the probability distribution of measurement results, quantum statistical inference deals with the inverse problem of estimating quantities related to the preparation of the quantum system, based on the measurement data.

The first papers dealing with quantum statistical problems appeared in the seventies \cite{Helstrom69,Yuen&Lax,Yuen&Lax&Kennedy,Belavkin,Holevo} and tackled issues such as quantum Cram\' er-Rao bounds for unbiased estimators, optimal estimation for families of states possessing a group symmetry, estimation of Gaussian states, optimal discrimination between non-commuting states. In recent years there has been a renewed interest in the field \cite{Hayashi.editor, Hayashi.book,Paris.editor,Barndorff-Nielsen&Gill&Jupp} and the advances in quantum engineering have led to the first practical implementations of theoretical methods \cite{Mabuchi,Hannemann&Wunderlich,Smith&Silberfarb&Deutsch&Jessen}.
An illustrating example is that of quantum homodyne tomography
\cite{Vogel&Risken,D'Ariano.2,Leonhardt.Munroe}, a measurement technique developed in quantum optics, which allows the estimation with arbitrary precision
\cite{Gill&Guta&Artiles,Butucea&Guta&Artiles} of the state of a monochromatic beam of light, by repeatedly measuring a sufficiently large number of identically prepared beams \cite{Smithey,Breitenbach&Schiller&Mlynek,Zavatta}.

Asymptotic inference is now a well established topic in quantum statistics, with many
papers \cite{Massar&Popescu,Cirac,Vidal,Gill&Massar,Keyl&Werner,Bagan&Baig&Tapia,Hayashi&Matsumoto,Hayashi&Matsumoto2,Bagan&Gill,Gill} concentrating on the problem of estimating an unknown state $\rho$ using the results of measurements performed on
$n$ quantum systems, identically prepared in the state $\rho$. For two dimensional systems, or qubits, the optimal state estimation problem has an explicit solution \cite{Bagan&Gill} in the special context of Bayesian inference, with invariant priors and figure of merit (risk) based on the fidelity distance between states. However this particular optimization method does not work for more general priors or loss functions and it seems to be limited to the qubit case. In the pointwise approach, Hayashi and Matsumoto \cite{Hayashi&Matsumoto} showed that the Holevo bound \cite{Holevo} for the variance of locally unbiased estimators can be attained asymptotically, and described a sequence of measurements achieving this purpose. Their results, building on earlier work \cite{Hayashi.japanese,Hayashi.conference}, provide the first evidence for the emergence of a Gaussian limit in the problem of optimal state estimation for qubits.

This paper together with the closely related works
\cite{Guta&Kahn,Guta&Janssens&Kahn} extend the results of Hayashi and
Matsumoto, and aim at developing quantum statistical analogues of fundamental concepts and tools in asymptotic statistics, such as convergence of statistical experiments and local asymptotic normality. The idea of approximating a sequence statistical models by a family of Gaussian distributions appeared in \cite{Wald}, and was fully developed by Le Cam \cite{LeCam} who introduced the term ``local asymptotic normality''. Among the many applications in mathematical statistics, local asymptotic normality is essential in asymptotic optimality theory and explains the asymptotic normality of certain estimators such as the maximum likelihood estimator. Based on the same principle, the paper \cite{Guta&Kahn} shows  that a similar phenomenon occurs in quantum statistics: the family of joint states of $n$ identically prepared  qubits converges to a family of Gaussian states of a quantum oscillator with unknown displacement. More precisely, there
exists a physical transformation (quantum channel) which maps the joint state of the spins into the oscillator state, such that local rotations around a fixed spin direction correspond to displacements of a thermal equilibrium state. In
\cite{Guta&Janssens&Kahn}, it was further shown that the passage to the limit can be physically implemented by transferring the joint qubits state to an approximate Gaussian state of a Bosonic field through a spontaneous emission coupling. After the transfer, the parameters of the initial qubit state can be estimated by means
of standard measurements in the field, which turns out to be optimal with respect to various criteria and a large class of loss functions.


In this paper we consider the general set-up of identically prepared finite dimensional quantum systems and prove a different version of the local asymptotic normality principle which we call weak convergence, in analogy with the classical statistics terminology. To motivate the result we build the first elements of a theory of weak convergence of quantum statistical models in close relation with the work of Petz on quantum sufficiency \cite{Petz86,Petz&Jencova,Ohya&Petz}. Our results add to the accumulating evidence for an underlying theory of quantum statistical experiments and quantum statistical decisions, which parallels the classical framework, but in the same time has
new `quantum' features generating a fruitful interaction between Mathematical Statistics, Quantum Information and Operator Algebras.

Before presenting the structure of the paper, here is a short summary of the key concept and ideas used in the paper. By adopting the terminology introduced by Le Cam \cite{LeCam} we call a quantum statistical experiment a family
$$
\mathcal{E} := (\mathcal{A}, \varphi_{\theta} :\theta\in\Theta),
$$
of states $\varphi_{\theta}$ on a von Neumann algebra $\mathcal{A}$ indexed by a parameter set $\Theta$. One may think of the quantum system as the carrier of a type of statistical information about the unknown parameter $\theta$ encoded by Nature (or an
adversary) in the state $\varphi_{\theta}$. Quantum decision problems such as state estimation or hypothesis testing can be formulated as a game between Nature who has the choice between different parameters $\theta$ and the physicist who tries to extract the maximum amount of information about the chosen $\theta$ for a given statistical purpose.

Quantum sufficiency deals with the situation when two such experiments
$$
\mathcal{E} := (\mathcal{A}, \varphi_{\theta} :\theta\in\Theta), \qquad
\mathcal{F} := (\mathcal{B}, \sigma_{\theta} :\theta\in\Theta),
$$
can be mapped into each other by quantum channels, i.e. there exist unit
preserving completely positive maps $T:\mathcal{A}\to\mathcal{B}$ and $S:\mathcal{B}\to\mathcal{A}$ such that
$$
\varphi_{\theta} =
\sigma_{\theta}\circ T, \qquad \sigma_{\theta} = \varphi_{\theta}\circ S, \quad \forall \theta.
$$
In this case it is clear that the two experiments are equivalent from a statistical point of view and the solution to any decision problem concerning one experiment can be
easily mapped to the other.

What if we have two experiments which are not equivalent but are `close to each other'
in a statistical sense? In Section \ref{sec.quantum.sufficiency} we enlarge the concept of sufficiency by defining the notion of convergence of experiments whereby a sequence
$\mathcal{E}_{n}$ approaches asymptotically a limit experiment $\mathcal{E}$
$$
\mathcal{E}_{n} \to \mathcal{E} , \qquad n \to \infty .
$$
When convergence holds, statistical problems concerning the experiment
$\mathcal{E}_{n}$ can be cast into problems concerning the potentially simpler experiment $\mathcal{E}$ with vanishingly small loss of optimality for large $n$. An important example is that of local asymptotic normality which means roughly the following: the sequence $\mathcal{E}_{n}$ of experiments consisting of joint states $\varphi^{n}$ of $n$ identical quantum systems prepared independently in the same state $\varphi$, converges  to a limit experiment $\mathcal{E}$ which is described by a family of Gaussian states on an algebra of canonical commutation relations.

This paper is intended to be a self-contained introduction to the theory of quantum
statistical experiments and local asymptotic normality. In Section \ref{sec.lan} we give an account of the classical concepts which will later be extended to the
quantum domain. Sufficiency and equivalence of statistical experiments are defined in Section \ref{subsec.stat.exp}. We then show how equivalence classes of experiments
can be described using the notion of canonical measure and Hellinger transform
(see Section \ref{sec.canonical.exp}). This enables us to define weak convergence of experiments as the pointwise convergence of the Hellinger transforms for all finite subsets of the parameter space. In parallel with the weak convergence we introduce the stronger topology of the Le Cam distance between two experiments.
This distance is based on the existence of a randomization mapping  the first experiment as close as possible to the second, and the other way around (see Section \ref{subsec.convergence.exp}).  We close the exposition of the classical theory with the exact formulation of local asymptotic normality. Given a ``smooth'' $m$-dimensional
family of distributions $P_{\theta}$ with $\theta\in\Theta\subset\mathbb{R}^{m}$ we
consider the experiments $\mathcal{E}_{n}$ consisting of $n$ independent, identically distributed variables $X_{1},\dots ,X_{n}$ with distribution $P_{\theta}$ where
$\theta :=\theta_{0}+u/\sqrt{n}$ lies in a local neighborhood of a fixed point
$\theta_{0}$, parametrized by $u$.
Then $\mathcal{E}_{n}$ converges weakly to a Gaussian shift experiment consisting of a single $m$-dimensional normal variable with distribution
$N(u, I_{\theta_{0}}^{-1})$ having unknown center $u$ and variance equal to the inverse of the Fisher information of $P_{\theta}$ at $\theta_{0}$ (see Section \ref{subsec.lan}).

Section \ref{sec.quantum.sufficiency} begins with a brief review of quantum sufficiency followed by the characterization of equivalence classes of experiments through the
canonical state (cf. Theorem \ref{th.canonical.state}). The latter gives the expectation of monomials of Connes cocycles $[D\varphi_{\theta}, D\varphi]_{t}$ for arbitrary
$\theta\in\Theta$ and $t\in\mathbb{R}$, and plays a similar role to that of the Hellinger transform of the classical case. Section \ref{sec.weak.strong} deals with the relation between weak and strong convergence of experiments. We show that for finite parameter sets $\Theta$ the weak and strong topologies coincide, under certain assumptions. The quantum Central Limit Theorem which is presented in Section \ref{sec.qclt} is one of the main ingredients of our result.

Finally, in Section \ref{sec.qlan} we prove the quantum local asymptotic normality Theorem \ref{th.qlan} as {\it weak} convergence of the i.i.d. experiment $\varphi_{\theta_{0}+u/\sqrt{n}}^{n}$ to a quantum Gaussian shift experiment $\phi_{u}$, which is the main result of the paper. This theorem holds for smooth families of  states on matrix algebras of arbitrary finite dimension, and it is complementary to the result of \cite{Guta&Kahn} concerning {\it strong} convergence for qubit states. For pedagogical reasons we first prove the result for a unitary family of states
in Section \ref{sec.unitary.one.parameter.lan}, which could be seen as a purely quantum experiment, after which we allow the change in eigenvalues leading to the presence of a classical Gaussian component in the limit experiment.

\section{Classical statistical experiments}\label{sec.lan}

In this section we describe the notion of local asymptotic normality and its significance
in statistics \cite{LeCam,Torgersen,Strasser,vanderVaart}. Suppose that we observe a sample $X_{1}, \dots ,X_{n}$ with $X_{i}$ taking values in a measurable space $(\Omega, \Sigma)$ and assume that $X_{i}$ are independent, identically distributed with
distribution $P_{\theta}$ indexed by a parameter $\theta$ belonging to an open subset $\Theta\subset\mathbb{R}^{m}$.
The full sample is a single observation from the product $P_{\theta}^{n}$ of
$n$ copies of $P_{\theta}$ on the sample space $(\Omega^{n}, \Sigma^{n})$.
The family of probability distributions $\left( P_{\theta}^{n} : \theta\in\Theta\right)$ is called a {\it statistical experiment} and the point of local asymptotic normality is to show that for large $n$ such statistical experiments can be approximated by Gaussian experiments after a suitable reparametrization. Let us fix a value $\theta_{0}$, define a local parameter
$u=\sqrt{n} (\theta-\theta_{0})$ and rewrite $P_{\theta}^{n}$ as
$P_{\theta_{0}+ u/\sqrt{n}}^{n}$ seen as a distribution depending on the
parameter $u$. We will show that for large $n$ the experiments
$$
\left(P_{\theta_{0} + u/\sqrt{n}} : u \in \mathbb{R}^{m}\right)
\qquad {\rm and} \qquad
\left( N( u, I_{\theta_{0}}^{-1}) : u \in \mathbb{R}^{m}\right),
$$
have similar statistical properties for ``smooth'' models $\theta\mapsto P_{\theta}$.
The point of this result is that while the original experiment may be difficult to analyze, the limit one is a tractable {\it Gaussian shift} experiment which can give us information about the original one, for instance in the form of lower bounds of estimation errors.  Let $p_{\theta}$ be the density of $P_{\theta}$ with respect to some
measure $\mu$. In the second experiment we observe a single sample from the normal distribution with unknown mean $u$ and fixed variance $I_{\theta_{0}}^{-1}$, where
$$
\left[I_{\theta_{0}}\right]_{ij} =
\mathbb{E}_{\theta_{0}}
\left[ \dot{\ell}_{\theta_{0}, i} \dot{\ell}_{\theta_{0}, j}
 \right],
$$
is the Fisher information matrix at $\theta_{0}$, with $\dot{\ell}_{\theta,i} := \partial\log p_{\theta}/\partial\theta_{i}$.


In the following subsections we will introduce the key concepts needed to understand local asymptotic normality: sufficiency, statistical equivalence, canonical measure, convergence of experiments.

\subsection{Statistical experiments, sufficiency, randomizations}
\label{subsec.stat.exp}

A typical statistical problem can be formulated as follows: given a sample
$X$ from a distribution $P_{\theta}$ over the measure space $(\Omega, \Sigma)$, find  $\hat{\theta}$ depending o $X$, an estimator of the unknown parameter $\theta\in\Theta$ such that the expected value of the distance $d(\theta, \hat{\theta})$ is small.
In general the space $\Theta$ need not be finite dimensional, for instance in the
case of estimating an unknown probability density on $\mathbb{R}$.

The estimation problem is an example of a {\it statistical decision problem}, a broad framework containing estimation as well as hypothesis testing problems. Clearly it is important to understand how much `statistical information' is contained in the experiment
$\mathcal{E} := ( P_{\theta} : \theta\in \Theta)$, when is an experiment more informative than another, and when
two experiments are close to each other in a statistical sense. Such questions have
been the main motivation for the development of the theory of statistical experiments pioneered by Le Cam \cite{LeCam}. In this section we will present some basic ideas of this theory, the converging point being the notion of local asymptotic normality.
For more information we refer to the monographs
\cite{LeCam,Torgersen,Strasser,vanderVaart}.


Let us start by explaining the notion of  {\it sufficiency} at the hand of an example. Let $X_{1},\dots , X_{n}$ be independent identically distributed random variables with values in
$\{0,1\}$ and distribution $P_{\theta} := ( 1-\theta,\theta)$ with $\theta\in (0,1)$,
and  denote  $\mathcal{E}_{n}:=( P_{\theta}^{n} : \theta\in\Theta)$ as before.
It is easy to see that $\bar{X}_{n} = \frac{1}{n} \sum_{i=1}^{n} X_{i}$ is an
unbiased estimator of $\theta$ and moreover it is a {\it sufficient statistic} for
$\mathcal{E}_{n}$, i.e. the conditional distribution $P_{\theta}^{n}(\cdot | \bar{X}_{n} = \bar{x})$ does not depend on $\theta$! In other words the dependence on $\theta$ of the total sample $(X_{1},X_{2},\dots , X_{n})$ is completely captured by the statistic
$\bar{X}_{n}$ which can be used as such for any statistical decision problem concerning $\mathcal{E}_{n}$. If we denote by $\bar{P}^{(n)}_{\theta}$ the distribution of $\bar{X}_{n}$ then the experiment
$\bar{\mathcal{E}}_{n} = ( \bar{P}^{(n)}_{\theta} : \theta\in\Theta)$ is statistically equivalent to $\mathcal{E}_{n}$. To convince ourselves that $\bar{X}_{n}$ does contain the same statistical information as $(X_{1},\dots , X_{n})$, we show that we can
simulate the latter by using a sample from $\bar{X}_{n}$ and an additional random
variable $Y$ uniformly distributed on $[0,1]$. Indeed for every fixed value $\bar{x}$ of
$\bar{X}_{n}$ there exists a measurable function
$$
f_{\bar{x}}: [0,1] \to \{0,1 \}^{n},
$$
such that  the distribution of $f_{\bar{x}} (Y)$ is $P^{n}_{\theta}( \cdot| \bar{X}_{n} = \bar{x})$ or
$$
\lambda( f_{\bar{x}}^{-1} (x_{1} ,\dots , x_{n}) ) =
P^{n}_{\theta}(  x_{1}, \dots ,x_{n} | \bar{X}_{n} = \bar{x}),
$$
where $\lambda$ is the Lebesgue measure on $[0,1]$. Then
$$
F(\bar{X}_{n}, Y) := f_{\bar{X}_{n}} (Y),
$$
has distribution $P_{\theta}^{n}$. The function $F$ is an example or
{\it randomized statistic} and it is a particular case of a more general construction
called {\it randomization} which should be seen as a transformation of an experiment into another which typically contains less information than the original one. We will give a short account of this notion in the case of dominated experiments. An experiment $\mathcal{E} =(P_{\theta} : \theta\in \Theta)$
on $(\Omega, \Sigma)$ is called {\it dominated} if there exists a $\sigma$-measure
$\mu$ such that $P_{\theta}\ll \mu$ for all $\theta$. We will often use the notation $P_{\theta}\sim \mu$ meaning that for any $A\in \Sigma$, $\mu(A) =0$ if and only if
$P_{\theta}(A)=0$ for all $\theta$.
\begin{definition}\label{def.stochoperator}
A positive linear map
$$M_{*}: L^{1} (\Omega_{1}, \Sigma_{1} , \mu_{1})\to L^{1}(\Omega_{2}, \Sigma_{2} , \mu_{2})$$
 is called a {\it stochastic operator} or {\it transition} if $\| M_{* } (g)\|_{1} = \|g \|_{1}$ for every $g\in L_{+}^{1}(\Omega_{1})$.
\end{definition}
\begin{definition}\label{def.markovop}
A positive linear map
$$M: L^{\infty} (\Omega_{2}, \Sigma_{2} , \mu_{2})\to L^{\infty}(\Omega_{1}, \Sigma_{1} , \mu_{1})$$
 is called a {\it Markov operator} if
$M \mathbf{1} =\mathbf{1}$, and if for any $f_{n}\downarrow 0$ in $L^{\infty}(\Omega_{2}) $ we have $M f_{n}\downarrow 0$.
\end{definition}
The pair $(M, M_{*})$ with $M$ and $M_{*}$ as above is called a dual pair if
$$
\int f M (g) d\mu_{1} = \int M_{*}(f) g d\mu_{2},
$$
for all $f\in L^{1}(\Omega_{1})$ and $g\in L^{\infty}(\Omega_{2})$. It is a theorem that for any stochastic operator $M_{*}$ there exists a unique dual Markov operator $M$ and
conversely, for any Markov operator $M$ there exists a unique dual stochastic operator $M_{*}$.
\begin{definition}\label{def.randomization}
Let $\mathcal{E}_{i}= ( P^{\theta}_{i}: \theta\in \Theta )$
be two dominated statistical experiments on $(\Omega_{i},\Sigma_{i})$
with $\mathcal{P}_{i}\sim \mu_{i}$,  $i=1,2$. Then $\mathcal{E}_{2}$ is a {\it randomization} of $\mathcal{E}_{1}$ if any of the following equivalent conditions is satisfied:
\begin{enumerate}
\item[(i)]
thererat exists a stochastic operator $M_{*} : L_{1} (\Omega_{1}, \Sigma_{1} , \mu_{1}) \to L_{1} (\Omega_{2}, \Sigma_{2} , \mu_{2})$ such that
$$
M_{*}( dP^{\theta}_{1}/d\mu_{1}) = dP^{\theta}_{2}/d\mu_{2}, \qquad \forall\theta\, ;
$$.
\item[(ii)]
there exists a Markov operator $M: L^{\infty} (\Omega_{2}, \Sigma_{2} , \mu_{2})\to
L^{\infty} (\Omega_{1}, \Sigma_{1} , \mu_{1})$ such that
$$
P^{\theta}_{2} = P^{\theta}_{1}\circ M, \qquad \forall\theta.
$$
\end{enumerate}
\end{definition}
A statistic $f:\Omega_{1}\to \Omega_{2}$ generates a sub$-\sigma-$field $\Sigma_{0}\subset \Sigma$ and a randomization which is the restriction of the measures $P^{\theta}$ to $\Sigma_{0}$. At the level of Markov operator this is simply described by the embedding of $L^{\infty}(\Omega, \Sigma_{0}, \mu)$ into
$L^{\infty}(\Omega, \Sigma, \mu)$.


In general by passing to a sub$-\sigma-$field some information about the initial distribution is lost. It turns out that the concept of randomization is the proper generalization of sufficiency. Indeed the next theorem shows that $\Sigma_{0}$ is sufficient for a dominated experiment $\mathcal{E}$ if this can be recovered by a randomization from the restricted experiment $\mathcal{E}_{0}$.
\begin{theorem}\label{th.sufficiency&randomizations}
Let $\mathcal{E} =( P_{\theta} : \theta\in\Theta )$ be a dominated experiment on
$(\Omega, \Sigma)$ and $\Sigma_{0}\subset\Sigma$ a sub-$\sigma$-field.
Denote by $\mathcal{E}_{0}$ the restriction of $\mathcal{E}$to $\Sigma_{0}$. Then $\Sigma_{0}$ is sufficient for $\mathcal{E}$ if and only if $\mathcal{E}$ is a randomization of $\mathcal{E}_{0}$.
\end{theorem}
Although the concept of randomization does not have a such a direct statistical
meaning as that of  randomized statistic, it is a very useful functional analytic generalization of the later and it is important as a mathematical tool due to the compactness of the space of randomizations in a certain weak topology.
\begin{definition}\label{def.stat.equivalent}
Two dominated experiments
$( P^{\theta}_{i} : \theta\in \Theta )$, $i=1,2$ are
{\it statistically equivalent} if each one is a randomization of the other.
\end{definition}
The idea of statistical equivalence is that for any statistical decision problem
the two experiments will have matching statistical procedures with the same risks, and thus contain `the same information'.


Finally we mention another useful characterization of sufficiency known as the
{\it Factorization Theorem} \cite{Strasser} which later will be extended to the quantum case.
\begin{theorem}\label{th.sufficiency.factorization}
Let $\mathcal{E} =( P_{\theta} : \theta\in\Theta )$ be a dominated experiment on
$(\Omega, \Sigma)$ with $P_{\theta}\sim \mu$,
and let $\Sigma_{0}\subset\Sigma$ be a sub-$\sigma$-field. Then $\Sigma_{0}$ is sufficient for $\mathcal{E}$ if and only if there exist a measurable function $h$ and for each $\theta$ a $\Sigma_{0}$-measurable function $g_{\theta}$ such that
$$
\frac{dP_{\theta}}{d\mu} = g_{\theta} h,  \qquad \mu-{\rm almost~surely}.
$$
 \end{theorem}

\subsection{The canonical measure and the Hellinger transform}
\label{sec.canonical.exp}
An important example of a sufficient statistic for $( P_{\theta} : \theta\in \Theta)$
is the likelihood ratio process.
\begin{definition}\label{def.lik.ratio}
Let $(P_{\theta} : \theta \in \Theta)$ be an experiment over $(\Omega, \Sigma)$
and suppose that $P_{\theta}\ll P_{\theta_{0}}$
for some fixed $\theta_{0}\in\Theta$ and all $\theta\in\Theta$.
The associated {\it likelihood ratio process} based at $\theta_{0}$ is
$$
\Lambda_{\theta_{0}} = \left\{ \theta \mapsto \frac{dP_{\theta}}{dP_{\theta_{0}}} \right\}.
$$
\end{definition}

Note that the likelihood ratio process is a rather  `large' statistic which takes values in
$\mathbb{R}^{|\Theta|}$
$$
\Lambda_{\theta_{0}} :\omega \mapsto \left\{\theta \mapsto
\frac{dP_{\theta}}{dP_{\theta_{0}}}(\omega) \right\} ,\quad \omega\in \Omega.
$$
The choice of the base point $\theta_{0}$ is not important as long as the distributions
$P_{\theta}$ are dominated by $P_{\theta_{0}}$. A variation on this can be considered
if we restrict to a finite set $\Theta$ of parameters.
In this case there exists a `standard representation' of statistical experiments such that statistically equivalent experiments have the same representation. Let $\mathcal{E}= (P_{\theta} : \theta \in \Theta)$ on
$(\Omega, \Sigma )$ and define $\mu := \sum_{\theta\in \Theta} P_{\theta}$ which will play the role of $P_{\theta_{0}}$.
Then the vector of likelihood ratios $V:=(dP_{\tau}/d \mu)_{\tau\in \Theta}$ seen as a
$\mathbb{R}^{|\Theta|}-$valued random variable on $(\Omega, \Sigma)$ induces the law
$\sigma_{\mathcal E} = \mathcal{L}(V~|~ \mu)$ called the {\it canonical measure} of $\mathcal{E}$. Note that neither $\mu$ nor $\sigma_{\mathcal E}$ is a probability distribution, but they both have mass $|\Theta|$. The experiment consisting in observing $V$ is called the {\it canonical experiment} and has law $Q^{\theta}:=\mathcal{L}\left( V~|~ P^{\theta}\right)$.
Because the likelihood ratio process is sufficient for $\mathcal{E}$, the canonical experiment is statistically equivalent to $\mathcal{E}$ and the distribution
$Q^{\theta}$  is supported by the simplex
$$
S_{\Theta}:= \left\{v=(v_{\theta}) \in \mathbb{R}_{+}^{ |\Theta|}, ~\sum_{\theta} v_{\theta}=1\right\}.
$$
We can now write
$$
Q^{\theta}(B) = \mathbb{E}_{\theta} \mathbf{1}_{B} (V)  =
                         \mathbb{E}_{\mu} \mathbf{1}_{B} (V) \frac{dP^{\theta}}{d \mu}=
                         \mathbb{E}_{\mu} \mathbf{1}_{B} (V) V_{\theta} =
                         \int_{B} v_{\theta} \sigma_{\mathcal E}(dv),
$$
which implies that
$$
Q^{\theta} (dv) = v_{\theta} \sigma_{\mathcal E}(dv),
$$
and thus the canonical experiment over the fixed measure space $S_{\Theta}$ is uniquely determined by the canonical measure $\sigma_{\mathcal E}$. Note that not every measure on the simplex is the canonical measure of some experiment.
\begin{theorem}\label{th.canonical.measure}
Two statistical experiments with the same finite parameter space $\Theta$ are statistically equivalent if and only if their canonical measures coincide.
 \end{theorem}
The canonical measure is at its turn completely characterized by the
{\it Hellinger transform} which is the function
$\eta_{\mathcal{E}}: \mathcal{S}_{\Theta}\to\mathbb{R}$ given by
$$
z\mapsto \eta_{\mathcal{E}} (z) =
\int_{\mathcal{S}_{\Theta}} \, \prod_{\theta\in \Theta}
v_{\theta}^{z_{\theta}} \sigma_{\mathcal{E}}(dv).
$$
The Hellinger transform is a continuous function on the interior of
$\mathcal{S}_{\Theta}$ taking values in $[0,1]$.
Note that if $\Theta= \{1,2\}$ and if $z\in \mathcal{S}_\Theta$ is given by
$ z_{1} = z_2= 1/2$ then
$$
\eta_{\mathcal{E}}(1/2, 1/2) = \int_{\mathcal{S}_{\Theta}} \, \sqrt{v_{1}v_{2}} \sigma_{\mathcal{E}}(dv) = \int \sqrt{\frac{dP_{1}}{d \mu} \frac{dP_{2}}{d \mu}} d\mu,
$$
 which is the {\it affinity} of $P_{1}$ and $P_{2}$ appearing in the well known Hellinger distance
 $$
 h(P_{1}, P_{2}) =
 \int \left(\sqrt{\frac{dP_{1}}{d\mu}} - \sqrt{\frac{dP_{2}}{d\mu}} \right)^{2} d\mu
 = 2 (1 - \eta_{\mathcal{E}}( 1/2, 1/2 )).
 $$
 \subsection{Convergence of statistical experiments}\label{subsec.convergence.exp}

How can we compare two statistical experiments
$\mathcal{E}_{i} =\left(P^{(i)}_{\theta} : \theta \in \Theta\right)$  on two different measure spaces
$(\Omega_{i}, \Sigma_{i})$ for $i=1,2$ ? When can we say that one is more informative then the other, or that the two are very close to each other ? More specifically we will be interested in the situation where a sequence of experiments $\mathcal{E}_{n}$ converges to a fixed one $\mathcal{E}$. A natural route is to compare their canonical measures.
\begin{definition}\label{def.weak.convergence.classical}
We say that a sequence of experiments $\mathcal{E}_{n}:=\left( P^{(n)}_{\theta} : \theta\in\Theta\right)$ converges weakly to an experiment
$\mathcal{E}:=\left( P_{\theta} : \theta\in\Theta\right)$ if for every finite
$ \mathcal{I}\in \Theta$, the sequence of canonical measures of $\mathcal{E}_{n}$ converges weakly (in distribution) to the canonical measure of $\mathcal{E}$.
\end{definition}
Another possibility is to compare the likelihood ratio processes
$$
\Lambda^{(n)}_{\theta_{0}} =
\left\{ \theta \mapsto \frac{dP^{(n)}_{\theta}}{dP^{(n)}_{\theta_{0}}}
\right\} \qquad  {\rm and} \qquad
\Lambda_{\theta_{0}} = \left\{ \theta \mapsto \frac{dP_{\theta}}{dP_{\theta_{0}}}
\right\},
$$
by demanding convergence in distribution of the marginals of these processes
for all finite sets $\mathcal{I} \subset \Theta$.
\begin{theorem}\label{th.weak.convergence}
Let $\mathcal{E}$ be such that $P_{\theta}\ll P_{\theta_{0}}$ for all $\theta$.
Then the following are equivalent:
\begin{itemize}
\item[(i)]
The sequence $\mathcal{E}_{n}$ converges weakly to $\mathcal{E}$.
\item[(ii)]
For any finite subset $\mathcal{I}\subset \Theta$, the sequence of Hellinger transforms $\eta_{\left. \mathcal{E}_{n}\right|\mathcal{I}}$ converge  to
$\eta_{\left. \mathcal{E}\right|\mathcal{I}}$
pointwise on $\mathcal{S}_{\mathcal{I}}$.
\item[(iii)]
The sequence of likelihood ratio processes $\Lambda^{(n)}_{\theta_{0}}$
converges to $\Lambda_{\theta_{0}}$ marginally in distribution.
\end{itemize}
\end{theorem}
\begin{ex}{\rm Consider a binomial variable with parameters $n$ and success
probability $\theta/n $:
$P^{(n)}_{\theta}(k) = \binom{n}{k}(\theta/n )^{k}(1-\theta/n)^{n-k}$, and the corresponding experiment $\mathcal{E}_{n}$ with $\theta$ ranging over the finite set $\{\theta_{1} , \dots , \theta_{p}\}$.
Then the Hellinger transform is
$$
\eta_{\mathcal{E}_{n}} (v_{1}, \dots , v_{p})= \left( \prod_{i=1}^{p} \left(\frac{\theta_{i}}{n}\right)^{v_{i}} + \prod_{i=1}^{p} \left(1- \frac{\theta_{i}}{n}\right)^{v_{i}} \right)^{n}.
$$
As $n\to\infty$ this converges pointwise to
$$
\eta(v_{1}, \dots , v_{p}) = \exp\left(\prod_{i=1}^{p} \theta_{i}^{v_{i}} -\sum_{i=1}^{p} \theta_{i}v_{i} \right),
$$
which is the Hellinger transform of an experiment consisting of observing a
Poisson variable with mean belonging to the set $\{\theta_{1}, \dots , \theta_{p}\}$.
}
\end{ex}

\begin{ex}{\rm The central example of this paper is that of local asymptotic normality.
Let $\mathcal{E}_{n}$ be the experiment consisting in observing a sample
$X_{1}, \dots X_{n}$ of independent identically distributed random variables with distribution $P_{\theta_{0} + u/\sqrt{n}}$, where $u\in \mathbb{R}^{m}$ should be seen as the unknown {\it local parameter} and we assume sufficient ``smoothness'' for the map
$\theta\mapsto P_{\theta}$. The claim is that
$$
\mathcal{E}_{n}:= \left( P_{\theta_{0} + u/\sqrt{n}} : u\in \mathbb{R}^{m}\right) \longrightarrow \left( N( u, I_{\theta_{0}}^{-1}) : u\in \mathbb{R}^{m}\right)
$$
where in the limit experiment we observe a single sample from the normal distribution with unknown mean $ u$ and fixed variance $I_{\theta_{0}}^{-1}$.
This claim will be detailed in Section \ref{subsec.lan}.
}
\end{ex}
Although minimalist with respect to the set of required relations, the concept of weak convergence is sufficiently strong to allow the derivation of certain statistical properties
of the sequence $\mathcal{E}_{n}$ from those of the limit experiment $\mathcal{E}$. A stronger convergence concept is that introduced by Le Cam using randomizations. As shown in Section \ref{subsec.stat.exp}, we can check statistical equivalence of two experiments by finding randomizations which map on experiment into the other. Naturally, when this can be done only approximately we think of the two experiments as being close to each other.
\begin{definition}\label{def.deficiency}
Let $\mathcal{E}_{i} := (P^{\theta}_{i} : \theta\in\Theta)$ be two statistical experiments dominated by $\mu_{i}$ for $i=1,2$. The {\it deficiency} of $\mathcal{E}_{1}$ with respect to $\mathcal{E}_{2}$ is the quantity
$$
\delta(\mathcal{E}_{1}, \mathcal{E}_{2}) := \inf_{M}\sup_{\theta}
\| P^{\theta}_{1}\circ M - P^{\theta}_{2}\|,
$$
where the infimum is taken over all Markov operators
$$
M: L^{\infty} (\Omega_{2}, \Sigma_{2}, \mu_{2})\to
     L^{\infty} (\Omega_{1}, \Sigma_{1}, \mu_{1}),
$$
and $\|\cdot\|$ is the total variation norm.
The {\it Le Cam distance} between $\mathcal{E}_{1}$ and $\mathcal{E}_{2}$ is defined as
$$
\Delta(\mathcal{E}_{1}, \mathcal{E}_{2}) = \mathrm{max}\left\{
\delta(\mathcal{E}_{1}, \mathcal{E}_{2}), \delta(\mathcal{E}_{2}, \mathcal{E}_{1})
\right\}.
$$
\end{definition}
We remind the reader that the total variation norm can be written in terms of the $L^{1}-$norm distance between the probability densities
$$
 \| P^{\theta}_{1}\circ M - P^{\theta}_{2}\|=
\frac{1}{2}\left\| M_{*}\left(\frac{dP^{\theta}_{1}}{d\mu_{1}} \right) -  \frac{dP^{\theta}_{2}}{d\mu_{2}}\right\|_{1}.
$$
The deficiency measure satisfies the triangle inequality
$\delta(\mathcal{E}, \mathcal{F}) + \delta(\mathcal{F}, \mathcal{G})\geq \delta(\mathcal{E}, \mathcal{G})$ but is not symmetric. This is remedied by the Le Cam distance which is a mathematical semi-distance. It can be shown that two experiments are at distance zero from each other if and only if they are statistically equivalent in the sense of Definition \ref{def.stat.equivalent}, and thus $\Delta$ defines a proper distance on the space of equivalence classes of experiments.

The relation between the strong convergence in the Le Cam distance and the weak convergence in the sense of convergence of canonical measures is given by the following theorem.
\begin{theorem}\label{th.strong.equivalent.weak.finite}
Let $\Theta$ be a finite set. Then strong convergence of experiments in the sense of Le Cam is equivalent to weak convergence of the canonical measures.
\end{theorem}
If $\Theta$ is not finite then weak convergence implies strong convergence
under the additional uniformity assumption: for any $\epsilon>0$ there exists a finite set
$I\subset\Theta$ such that
$$
\limsup_{n\to\infty} \sup_{\theta} \inf_{\tau\in I} \| P^{(n)}_{\theta} - P^{(n)}_{\tau} \| <\epsilon.
$$

Although the Le Cam distance is very appealing from the mathematical point of view, it is often difficult to calculate and will not play any role in our discussion. However, in a quantum theory of experiments the Le Cam distance should play a central role and some encouraging results in this direction exist already. In
\cite{Guta&Kahn,Guta&Janssens&Kahn} it is shown that the quantum version of the local asymptotic normality with the Le Cam type convergence holds for identically prepared qubits with the limit experiment being a family of displaced thermal equilibrium states. In \cite{Guta&Matsumoto}, the problem of optimal
cloning of mixed quantum Gaussian states is solved along lines similar to the solution
of the classical problem of finding the deficiency between two Gaussian shift experiments.

\subsection{Local asymptotic normality}\label{subsec.lan}

We return now to the second example of Section \ref{subsec.convergence.exp}.
A sufficient smoothness property for the family $ (P_{\theta} : \theta \in \Theta)$ is
the differentiability of $\theta\mapsto \sqrt{p_{\theta}}$ in quadratic mean:
there exists a vector of measurable functions $\dot{\ell}_{\theta}=(\dot{\ell}_{\theta,1} , \dots , \dot{\ell}_{\theta,k} )^{T}$ such that
$$
\int \left[ \sqrt{p_{\theta+u} } - \sqrt{p_{\theta}} - \frac{1}{2} u^{T} \dot{\ell}_{\theta} \sqrt{p_{\theta}} \right]^{2} d\mu = o(\|u\|^{2}).
$$
This condition is satisfied in many models and it is sufficient to have $\sqrt{p_{\theta}}(x)$ continuously differentiable in $\theta$ for almost all $x$ and the Fisher information  $I_{\theta}$ continuous in $\theta$.
\begin{theorem}\cite{vanderVaart}
Suppose that $\Theta$ is an open set in $\mathbb{R}^{m}$ and that the family
$(P_{\theta} : \theta\in\Theta)$ is differentiable in quadratic mean at $\theta_{0}$. Then
$$
\log \prod_{i=1}^{n} \frac{p_{\theta_{0}+u/\sqrt{n}}}{p_{\theta_{0}}}(X_{i}) =
\frac{1}{\sqrt{n}} \sum_{i=1}^{n} u^{T} \dot{\ell}_{\theta_{0}} (X_{i}) - \frac{1}{2n}u^{T}  I_{\theta_{0}} u
+ o_{P_{\theta_{0}}}(1).
$$
\end{theorem}

We refer to \cite{vanderVaart} for the proof of the theorem and outline here only the key points under the stronger assumption that $\ell_{\theta} (x) = \log p_{\theta}(x)$ is
twice differentiable with respect to $\theta$ for every $x\in \Omega$. Assume for simplicity that $\theta$ is a one dimensional parameter, then we have the expansion
$$
\log\prod_{i=1}^{n} \frac{p_{ \theta_{0} + u/\sqrt{n} } }{p_{\theta_{0}} } (X_{i})=
\frac{u}{\sqrt{n} } \sum_{i=1}^{n} \dot{\ell}_{\theta_{0}} (X_{i}) + \frac{1}{2} \frac{u^{2}}{n}
\sum_{i=1}^{n} \ddot{\ell}_{\theta_{0}} (X_{i}) + {\rm Rem}_{n}.
$$
The first term on the right side has mean zero because $P_{\theta}\ell_{\theta}=0$ and thus it can be written as $u\Delta_{n,\theta_{0}}$with $\Delta_{n,\theta_{0}}$
converging to a normal distribution of zero mean and variance $I_{\theta_{0}}$ by the Central Limit Theorem. The second term converges to $-\frac{1}{2}u^{2}I_{\theta_{0}}$ by the Law of Large Numbers. Thus we have the convergence in distribution for
$X\sim N(0, I_{\theta_{0}})$
$$
\log\prod_{i=1}^{n}  \frac{p_{ \theta_{0} + u/\sqrt{n} } }{p_{\theta_{0}} }(X_{i}) \to
u X  - \frac{1}{2} u^{2}I_{\theta_{0}} =
\log\frac{dN(uI_{\theta_{0}}, I_{\theta_{0}})}{ dN(0 , I_{\theta_{0}})}(X).
$$
\begin{theorem}
Let $\mathcal{E}_{n}:=(P^{n}_{\theta_{0} + u/\sqrt{n}} : u \in \mathbb{R}^{m} )$
be a sequence of experiments satisfying local asymptotic normality and
$\mathcal{E} =(N(u , I_{\theta_{0}}^{-1})  : u\in \mathbb{R}^{m})$. Then
$$
\mathcal{E}_{n} \to \mathcal{E}, \qquad n\to\infty,
$$
in the sense of weak convergence of experiments.
\end{theorem}

\section{Quantum statistical experiments}\label{sec.quantum.sufficiency}

The first steps in developing a quantum analogue of the classical theory of statistical experiments were taken by Petz \cite{Petz86}, and the latest results on quantum sufficiency can be found in \cite{Petz&Jencova}. We begin this section with the basic notions of quantum sufficiency. Later we will further extend the theory to cover approximate sufficiency through the notion of convergence of quantum statistical experiments. For a review of the complementary theory of quantum statistical inference we refer to \cite{Barndorff-Nielsen&Gill&Jupp}.

We remind the reader that a quantum mechanical system is modeled by a $C^*$-algebra $\mathcal A$, where the observables of the system  correspond to self-adjoint elements and the states are represented by normalized positive functionals on $\mathcal A$.
Let $\mathcal S=(\varphi_{\theta} :\theta\in\Theta)$ be a parametrized family
of states on $\mathcal A$, then the couple $\mathcal E=(\mathcal A,\mathcal S)$ is called a {\em  quantum statistical experiment}. We will mostly assume that $\mathcal A$ is also a von Neumann algebra, in which case the states $\varphi_{\theta}$
are required to be normal. Von Neumann algebras are the non-commutative analogues of classical algebras of bounded random variables $L^{\infty}(\Omega, \Sigma, \mu)$, and the normal states are the analogue of the probability distributions which are continuous with respect to $\mu$, i.e. their densities span the space $L^{1}(\Omega, \Sigma, \mu)$.

The interest in considering subsets of the whole set of
states is that in this way we can encode prior information about the preparation,
 for instance if we know that the state is pure, or that it has a block diagonal form.

Let $\mathcal B$ be another von Neumann algebra and let
$\alpha:\ \mathcal B\to \mathcal A$
be a linear map. Then $\alpha$ is a {\it channel}
if it is completely positive, unit-preserving and normal. Such maps are the
quantum versions of Markov operators (see Definition \ref{def.markovop}),
and their duals which act on states, are the quantum state transitions.
We will further suppose that all the channels are faithful, that is if
$\alpha(a)=0$ for some positive $a$ then $a=0$.

Let $\iE=(\mathcal A,  \varphi_{\theta}: \theta\in \Theta)$ be an experiment and
$\alpha:\ \mathcal B\to\mathcal A$  a channel.
The induced
experiment
$\iF=\iE\circ\alpha:=(\mathcal B, \varphi_{\theta}\circ\alpha: \theta\in
\Theta)$ is called a {\em randomization} of $\iE$. If also $\iE$ is a
randomization of $\iF$, i.e. there is a channel
$\beta:\ \mathcal A\to\mathcal B$, such that
$\varphi_{\theta}\circ\alpha\circ\beta=\varphi_{\theta}$ for all $\theta$, then the experiments
$\iE$ and $\iF$ are {\em statistically equivalent}. In this case, we also say that the channel
$\alpha$ is {\em sufficient for $\iE$}.
If $\mathcal B\subset \mathcal A$  is a subalgebra and the inclusion map
$\mathcal B\to\mathcal A$ is sufficient for $\iE$, then $\mathcal B$ is a
{\em sufficient subalgebra for $\iE$.} Note that a sufficient channel is intrinsically
related to the quantum experiment, in particular it may {\it not} be invertible on the whole set
of states of $\mathcal{A}$ as we will see in examples.


In order to give a characterization of quantum sufficiency, we first need  to describe its basic ingredients. We restrict to the case when all the states
in $\mathcal S$ are faithful, and we refer to \cite{Petz&Jencova} for the
more general situation. We denote the set of all such experiments with
parameter space $\Theta$ by $\iE(\Theta)$.
\begin{definition}\label{def.mod.group}
Let $\varphi$ be a state on $\mathcal{A}$. There exists a
unique group $\sigma^\varphi_t$ of automorphisms of $\mathcal A$ called
the modular group of $\varphi$ such that the following modular condition holds.  For each $a,b\in \mathcal A$, there is a function
$F\in \mathbb A(J)$, such that
$$
F(t)=\varphi(a\sigma^\varphi_t(b)),\quad F(t+i)=\varphi(\sigma^\varphi_t(b)a),
\qquad t\in\mathbb R,
$$
where  $\mathbb A(J)$ denotes the set of functions analytic in the strip
$$
J:=\{z\in \mathbb C,\  0<\mathrm{ Im}\, z< 1\},
$$
and continuous on
the closure $\bar J$.
\end{definition}
\begin{definition}\label{def.cocycle}
Let $\theta_0,\theta$ be two points in $\Theta$ and $\varphi:=\varphi_{\theta_0}$ and $\varphi_\theta$ be the corresponding states. The Connes cocycle derivative $u_{t}=[D\varphi_{\theta} , D\varphi]_t$ is a $\sigma$-strongly continuous one
parameter family of unitaries in $\mathcal A$ with the following properties
\cite{Takesaki2}:
\begin{enumerate}
\item[(a)] $u_{t}$ satisfies the cocycle condition
$u_{s} \,\sigma^\varphi_s (u_{t})=u_{t+s}$,\quad $s,t\in\mathbb R$.
\item[(b)] $u_t\sigma^\varphi_t(a)u_t^*=
\sigma^{\varphi_{\theta}}_t(a),\quad a\in\mathcal A, t\in \mathbb R.
$
\item[(c)] For all
$a,b\in \mathcal A$, there is a function $F\in \mathbb A(J)$,
such that
$$
F(t+i)=\varphi(au_t\sigma_t^\varphi(b)),
\quad F(t)=\varphi_{\theta}(u_t\sigma_t^\varphi(b)a),\qquad t\in
\mathbb R.
$$
\end{enumerate}
\end{definition}
The family of cocycle derivatives $([D\varphi_{\theta}, D\varphi]_{t}: t\in\mathbb{R}, \theta\in\Theta)$  is the quantum analogue of the likelihood ratio process (see Definition \ref{def.lik.ratio}). Indeed in the commutative case the modular group is
trivial and the above conditions are satisfied by
$u_{t} = \left(d P_{\theta}/dP_{\theta_{0}}\right)^{it}$.

In this paper we are particularly interested in the case of type I algebras $\mathcal{A}$ which appear more often in physical applications, i.e. matrix algebras
$M(\mathbb{C}^{d})$, the algebra $\mathcal{B}(\mathcal{H})$ for $\mathcal{H}$ separable infinite dimensional Hilbert space, and direct sums thereof. Then $\mathcal A$ admits a trace $\Tr$ and each state $\varphi$ is uniquely characterized by its  density operator $\rho$ as
$$
\varphi(a)=\Tr (\rho a),\qquad a\in \mathcal A.
$$
Let $\rho_\theta$ be the density operator for $\varphi_{\theta}$, then the modular group and the cocycle derivatives are given by
\begin{equation}\label{eq.modular.group.Connes.cocycle}
\sigma^\varphi_t(a)=\rho^{it}a\rho^{-it}\qquad  {\rm and} \qquad
[D\varphi_{\theta},D\varphi]_t=\rho_\theta^{it}\rho^{-it}.
\end{equation}

Note that if we put $a=b=1$ in (c) and if $F$ is the corresponding function in
$\mathbb A(J)$, then $F(i1/2)$ is the transition probability
$P_A(\varphi_{\theta},\varphi):= \mathrm{Tr}(\sqrt{\rho}\sqrt{\rho_{\theta}})$.
Moreover, for $p\in (0,1)$, we can define the relative quasi-entropy by
$$
S_p(\varphi_{\theta},\varphi)=\frac1{p(1-p)}(1-F(ip)) =\frac1{p(1-p)}(1-
\Tr(\rho^{p}\rho_\theta^{1-p})).
$$

Let $\mathcal A$ and $\mathcal B$ be
von Neumann algebras and let $\alpha:\mathcal B\to \mathcal A$ be a channel. Then the {\em multiplicative domain} of $\alpha$ is the
subalgebra $\mathcal B_\alpha\subset \mathcal B$,
 defined by
$$\mathcal B_\alpha:=\{a\in\mathcal B, \alpha(a^*a)=
\alpha(a)^*\alpha(a) \, :\,\ \alpha(aa^*)=\alpha(a)\alpha(a)^*\},
$$
and the restriction of $\alpha$ to the multiplicative domain is an isomorphism
onto $\alpha(\mathcal B_\alpha)$ if $\alpha$ is faithful.
\begin{theorem}\label{thm:suffch}\cite{Petz&Jencova}
Let $\mathcal E =(\mathcal{A}, \varphi_{\theta} :\theta\in\Theta)$ be a quantum statistical experiment and let $\varphi=\varphi_{\theta_0}$.
Let $\alpha:\mathcal B\to\mathcal A$ be a faithful channel, then the following are equivalent:
\begin{enumerate}
\item [(i)] $\alpha$ is sufficient for $\mathcal E$,
\item [(ii)]
$S_p(\varphi_{\theta},\varphi)=
S_p(\varphi_{\theta}\circ\alpha,\varphi\circ\alpha)$ for all $\theta$ and for some $p\in (0,1)$,
\item [(iii)] $[D\varphi_{\theta}, D\varphi]_t=
\alpha([D(\varphi_{\theta}\circ\alpha),
D(\varphi\circ\alpha)]_t)$ for all $\theta$ and $t\in \mathbb{R}$,
\item [(iv)] $\alpha(\mathcal B_\alpha)$ is a sufficient subalgebra for $\mathcal E$.
\end{enumerate}
\end{theorem}
Note that in the case that $\mathcal B$ is a subalgebra in $\mathcal A$, the
condition (iii) is equivalent to
\begin{enumerate}
\item[(iii').] $[D\varphi_{\theta},D\varphi]_t\in \mathcal B$ {\em for all
$\theta\in\Theta$ and $t\in\mathbb R$.}
\end{enumerate}

This implies that the subalgebra generated by the cocycle derivatives is
sufficient for $\iE$ and it is contained in any other sufficient
subalgebra, so that it is {\em minimal sufficient}. We will denote this
subalgebra by $\mathcal A_\iE$. Moreover, the cocycle
condition implies that $\mathcal A_\iE$ is invariant under the modular group
$\sigma_t^\varphi$.
For a channel $\alpha:\mathcal B\to\mathcal A$, the
conditions of the Theorem are equivalent to
the fact that the minimal sufficient subalgebra
$\mathcal B_\iF$ for the induced experiment $\iF=\iE\circ\alpha$ is contained in
the multiplicative domain of $\alpha$.
\begin{corollary}\label{cor:suffch}
Two statistical experiments $\mathcal{E}:=(\mathcal A , \varphi_\theta : \theta\in\Theta)$ and
$\mathcal{F}:=(\mathcal B , \sigma_\theta : \theta\in\Theta)$ are statistically equivalent if and
only if there exists an isomorphism $\alpha:\mathcal{B}_\mathcal{F} \to \mathcal{A}_\mathcal{E}$ between
their minimal sufficient algebras such that $\varphi_\theta\circ\alpha =\sigma_\theta$ for all $\theta$.
\end{corollary}
\begin{ex}\label{example}\rm Let  $\mathcal A=M_d(\mathbb C)$ and let
$\iE=(\mathcal A,\varphi_{\theta}: \theta\in \theta)$ be a quantum experiment.
Let $\mathcal A_0\subset \mathcal A$ be a subalgebra. Then there is a
decomposition
$$
\mathbb {C}^d=\bigoplus_{i=1}^m H_i^L\otimes H_i^R,
$$
with the
projections $p_i:\ \mathbb{C}^d\to H_i^L\otimes H_i^R$, such that
$\mathcal A_0$ is isomorphic to $\bigoplus_{i=1}^m B(H^L_i)\otimes 1_{H^R_i}$.
Let us also suppose that  $\mathcal A_0$ is invariant under $\sigma^\varphi_t$.
Then $\mathcal A_0$ is sufficient for $\mathcal E$ if and only if the density matrices have the form
\begin{equation}\label{eq:factor}
\rho_\theta=\sum_{i=1}^m\varphi_{\theta}(p_i) \rho^{L}_{\theta,i}\otimes \rho^{R}_i,\quad
\theta\in\Theta,
\end{equation}
where $\rho^{L}_{\theta,i}\in B(H^L_i)$,
$\rho^{R}_i\in B(H^R_i)$ are density matrices
(cf. \cite{Petz&Mosonyi}, see also \cite{Petz&Jencova} for an
infinite dimensional version). If $\mathcal A_0$ is the minimal sufficient
subalgebra, then the decomposition (\ref{eq:factor})
is the maximal decomposition obtained in
\cite{Koashi&Imoto}. Since any sufficient subalgebra contains the minimal
sufficient subalgebra, we may conclude that an arbitrary subalgebra
$\mathcal A_0$ is sufficient if and only if there is an orthogonal sequence of
projections  $\{p_i\}$  in  $\mathcal A_0$ with  $\sum_ip_i=\mathbf{1}$,
positive elements  $\rho_{\theta,i}\in\mathcal A_0$ and $\rho_i\in\mathcal A$
with supports $p_i$, commuting for all $\theta$, such that
$$
\rho_\theta=\sum_i\varphi_{\theta}(p_i)\rho_{\theta,i}\rho_i.
$$
This result is the quantum version
of the factorization Theorem \ref{th.sufficiency.factorization}.
\end{ex}

\subsection{Equivalence classes of experiments}

The notion of statistical equivalence of experiments as introduced in the previous Section
 defines an equivalence relation on $\iE(\Theta)$.  In this section, we want to describe the equivalence classes. The aim is to construct quantum analogues of the notions of canonical experiment and canonical measure described in
 Section \ref{sec.canonical.exp}.


Let $\mathcal E=(\mathcal A,\varphi_{\theta}:\theta\in\Theta)$ be an experiment in
$\mathcal E(\Theta)$.
Then the  equivalence class of $\mathcal E$ contains also  the restriction
$\mathcal E|_{\mathcal A_\mathcal E}$ to the minimal sufficient subalgebra
$\mathcal A_\iE$. We may therefore consider only experiments such that
$\mathcal A$ is generated by the cocycle derivatives. In
what follows $(\mathcal{A}_\mathcal E,H_\mathcal E,\xi_\iE)$ always
denotes the GNS representation of the minimal sufficient subalgebra  with respect to the state
$\varphi=\varphi_{\theta_{0}}$.

Let $G=G(\Theta)$ be the free group generated by the set of symbols
$$
\{ u_t(\theta)~:~ \ u_0(\theta)=u_t(\theta_0)=e, \,\theta\in\Theta, t\in\mathbb R\}.
$$
We denote by $L_1(G)$ the Banach space of all summable functions
$f:G\to \mathbb C$, with norm $\|f\|:=\sum_{g\in G}|f(g)|$. The dual space
$L_1(G)^*$ can be identified with the space $L_\infty(G)$ of bounded
functions over $G$, equipped with the supremum norm.

For each experiment $\mathcal E\in\mathcal E(\Theta)$ there is a unique  group
homomorphism
\begin{eqnarray*}
\pi_{\mathcal E}:\
G &\to& \mathcal U(H_\iE), \\
u_t(\theta)&\mapsto& [D\varphi_{\theta},D\varphi]_t,\quad \forall \theta\in\Theta, \, t\in\mathbb{R},
\end{eqnarray*}
thus $\pi_\iE$ is a unitary representation of $G$ on $H_\iE$.
We define a function on $G$ by
$$
\omega_\iE(g)=\<\xi_\iE,\pi_\iE(g)\xi_\iE\>=\varphi(\pi_\iE(g)),\quad g\in G.
$$
Then $\omega_\iE$ is a state, that is a positive definite function on $G$,
satisfying  $\omega_\iE(e)=1$ and will be called the {\em canonical state}
of the experiment  $\iE$. Since for any state $\omega$ we have
$|\omega(g)|\le \omega(e)=1$ for all $g\in G$, the set of all states is  a
subset in the unit ball of $L_\infty(G)$.   Clearly, the GNS
representation $\pi_{\omega_{\iE}}$ of $G$ with respect to
$\omega_\iE$ is equivalent with $\pi_\iE$.

From property (c) of the cocycle
derivatives we know that for any $\theta\in\Theta$ and $g\in G$ there is a
function
$F_{\mathcal{E},g,\theta}\in\mathbb A(J)$ such that
$$
F_{\iE,g,\theta}(t+i)=\varphi(\pi_\iE(g)[D\varphi_{\theta}, D\varphi]_t)=
\omega_\iE(gu_t(\theta)),
$$
and
$
|F_{\iE,g,\theta}(z)|\le 1
$
for all $z\in J$. We have the following characterization of the equivalence
classes of  experiments.

\begin{theorem}\label{th.canonical.state} Let $\mathcal E=(\mathcal A,\varphi_{\theta}:\theta\in\Theta)$ and
$\mathcal F=(\mathcal B,\psi_\theta :\theta\in \Theta)$ be experiments in $\mathcal E(\Theta)$
with $\mathcal A$ and $\mathcal B$ minimal sufficient. Then $\mathcal E$ is equivalent with
$\mathcal F$ if and only if $\omega_{\mathcal E}=\omega_{\mathcal F}$.
\end{theorem}


\noindent{\it Proof.}
Let $\mathcal E$ be equivalent with $\mathcal F$, then by Corollary
\ref{cor:suffch}, there is an
isomorphism $\alpha:\mathcal A\to  \mathcal B$, such
that $\varphi_{\theta}=\psi_\theta\circ\alpha$ and
$\alpha([D\varphi_{\theta} ,D\varphi]_t)=[D\psi_\theta,D\psi]_t$, $\theta\in\Theta$, $t\in\mathbb R$. (we remind the reader that
$\mathcal A_\iE=\mathcal A$ and $\mathcal B_\iF=\mathcal B$.)
By uniqueness of $\pi_\iF$, it follows that
$\pi_{\mathcal F}=\alpha\circ\pi_{\mathcal E}$ and
$$
\omega_{\mathcal F}=\psi\circ\pi_{\mathcal F}=\psi\circ\alpha\circ\pi_{\mathcal E}=\omega_{\mathcal E}.
$$
To prove the converse,  let $\omega_{\mathcal E}=\omega_{\mathcal F}=:\omega$,
then
$\pi_{\mathcal E}$ and $\pi_{\mathcal F}$ are equivalent, since they are both equivalent with $\pi_\omega$.
Hence there is a unitary $U:\ H_\iF\to H_\iE$, such that $\pi_\iF(g)=U^*\pi_\iE(g)U$ and the
cyclic vectors satisfy $U\xi_\iF$ = $\xi_\iE$. In particular $[D\psi_\theta,D\psi]_t=U^*[D\varphi_\theta,D\varphi]_tU$ and it is enough to prove that $\psi_\theta=\varphi_\theta\circ Ad_U$ for all $\theta\in\Theta$.
For $\theta\in \Theta$, $g\in G$,  the functions
$F_{\mathcal E,g,\theta}$ and $F_{\mathcal F,g,\theta}$ are in $\mathbb A(J)$
and coincide on $\mathbb R+i$, hence they coincide on $J$.
It follows that
$$
\psi_\theta(\pi_\iF(g))=
F_{\mathcal F,g,\theta}(0)=F_{\mathcal E,g,\theta}(0)=
\varphi_\theta(\pi_\iE(g))=
\varphi_\theta(U\pi_\iF(g)U^*),
$$
for all $g\in G$.
 Since the elements $\{ \pi_\iF(g),\ g\in G\}$
generate $\mathcal B$, the proof is finished.

\qed

\noindent{\bf Remark.}
Let us suppose that  $\iE$ is a binary experiment, that is, $\Theta$ consists
of two points $\{\theta_1,\theta_0\}$. Let $F$ be the analytic continuation of
the function
$t\mapsto \omega_\iE(u_t(\theta_1))$. Then the function
$$
\phi_\iE:\ (0,1)\ni p\to F(ip),
$$
can be viewed as
a quantum version of the Hellinger transform. If for some binary experiments
$\iE$ and $\iF$ we have
$\phi_\iE=\phi_\iF$, then clearly $\omega_\iE(u_t(\theta_1))=\omega_\iF(u_t(\theta_1))$ for all $t$, but, unlike the classical case, this is not enough to
 characterize quantum statistical equivalence, since we need the values of the
 canonical states on all products of $u_t(\theta_1)$. This corresponds to the
 results in \cite{Nagaoka&Ogawa}, where it is proved that, at least in finite
 dimensional case, quantum statistical equivalence cannot be determined by the
 class of quantum $f$-divergences, unless the experiments are commutative.

\subsection{The set of canonical states}
As we have seen, $\iE(\Theta)$ can be identified with a subset in the unit ball of
$L_\infty(G(\Theta))$ through the canonical state. In this section we will describe this subset.

For each $s\in \mathbb R$, we define an automorphism on $G$ as the extension of
the map
$$
\alpha_s(u_t(\theta))= u_s(\theta)^{-1}u_{t+s}(\theta),\quad \theta\in\Theta.
$$
Then $\alpha_s$, $s\in \mathbb R$ is a group of automorphisms on $G$.
If $\omega=\omega_\iE$ is a canonical state, then
the cocycle condition implies
$$
\pi_\iE(\alpha_s(u_t(\theta)))=[D\varphi_\theta,D\varphi]_s^*[D\varphi_\theta,D\varphi]_{t+s}=\sigma^\varphi_s(\pi_\iE(u_t(\theta))),
$$
It follows that
\begin{equation}\label{eq:KMS}
\pi_\iE(\alpha_s(g))=\sigma^\varphi_s(\pi_\iE(g)),\quad
g\in G,
\end{equation}
so that $\omega$ satisfies the modular condition with respect to $\alpha_s$.
Moreover, it follows from the properties of the Connes cocycle  that for
$g,h\in G$ and $\theta\in\Theta$, the  functions
$t\mapsto \omega(gu_t(\theta)\alpha_t(h))$ have an analytic continuation to
the strip $\tilde{J}\subset\mathbb{C}$ which is the reflection of $J$ with respect to the real axis (see Definition \ref{def.mod.group}), and they are
bounded by 1 in absolute value on $\tilde{J}$.
The next Theorem shows that this property completely
characterizes the canonical states.
\begin{theorem}\label{thm:canon}
Let $\omega$ be a state in $L_\infty(G)$. Then $\omega$ is the  canonical state for some experiment $\iE$ if and  only if  for each  $\theta\in\Theta$ and
$g,h\in G$, there is  a  function
$F_{g,h,\theta}\in \mathbb A(J)$, $|F_{g,h,\theta}(z)|\le 1$ for  $z\in J$,
satisfying
\begin{eqnarray*}
F_{g,h,\theta}(t+i)&=&\omega(gu_t(\theta)\alpha_t(h)),  \qquad t\in\mathbb R
\\
F_{g,h,\theta_0}(t)&=&\omega(\alpha_t(h)g),\quad g,h\in G,\
t\in \mathbb R,\quad
 F_{e,e,\theta}(0)=1,\quad \theta\in\Theta.
\end{eqnarray*}
\end{theorem}

\noindent{\it Proof.}
Note that the conditions for $\theta=\theta_0$ imply that $\omega$ satisfies
the modular condition for $\alpha_t$. If $\omega$ is a canonical state then by
Definition \ref{def.cocycle}, the function
$$
F_{g,h,\theta}(t)=\omega_\theta(u_t(\theta)\alpha_t(h)g),\qquad t\in\mathbb R
$$
satisfies the required conditions,
where
\begin{equation}\label{eq:theta}
\omega_\theta(g)=\varphi_\theta(\pi_\omega(g)), \qquad g\in G.
\end{equation}

For the converse, let $(\pi_\omega,H_\omega,\xi_\omega)$ be the GNS triple for $\omega$ and define $\mathcal M_\omega=\pi_\omega(G)''$. We will first show that
the state $\varphi=\<\xi_\omega,\,\cdot\, \xi_\omega\>$ is faithful
on $\mathcal M_\omega$.

Suppose that  $a$ is a positive element in $\mathcal M_\omega$,
such that $\varphi(a)=0$. Let $\mathbb C[G]$ be the algebra of all
finite complex-linear combinations of elements of $G$, then $\pi_\omega$
extends naturally to $\mathbb C[G]$ and $\pi_\omega(\mathbb C[G])$ is
a strongly dense *-subalgebra in $\mathcal M_\omega$.
By Kaplansky density theorem \cite{Kadison&Ringrose}, there is a net
$\{a_j\}_{j\in \mathcal I}$  of positive
elements in $\mathbb C[G]$,   such that
$\pi_\omega(a_j)$ converges strongly to $a^{1/2}$. By assumptions,
for  any $b,c\in \mathbb C[G]$ and $j\in \mathcal I$, there is a function
 $F_{j} := F_{a_{j}b^{*} ,c,\theta_{0}}\in \mathbb A(J)$, such that
$$
F_j(t+i)=\omega(a_jb^*\alpha_t(c)),\qquad F_j(t)=\omega(\alpha_t(c)a_jb^*).
$$
Since $\omega$ satisfies the modular condition, it is
is invariant under $\alpha_t$, so that  both $F_j(t)$ and
$F_j (i+t)$ converge uniformly on $\mathbb R$. By the maximum modulus principle,
$F_j(z)$ converges uniformly on $J$ to a function $F\in \mathbb A(J)$.
But since $|F_j(t+i)|^2\le \omega(a_jb^*ba_j)\omega(c^*c)\to 0$,
$F(t+i)=0$, for $t\in \mathbb R$ and  hence $F(z)=0$ on $\bar J$.
It follows that
$$
F(0)=\<\pi_\omega(c^*)\xi_\omega,
a^{1/2}\pi_\omega(b^*)\xi_\omega\>=0.
$$
As this is true for all $b,c\in \mathbb C[G]$, we get $a^{1/2}=0$.

Let now $U_t$ be the unitary on $H_\omega$,  given by
$U_t\pi_\omega(a)\xi_\omega=\pi_\omega(\alpha_t(a))\xi_\omega$,
$a\in \mathbb C[G]$ and let
$\sigma_t=Ad_{U_t}$. Then $\sigma_t\circ\pi_\omega=\pi_\omega\circ\alpha_t$
 and $\varphi$ satisfies the modular  condition for $\sigma_t$ on a
 $\sigma$-strongly  dense subset in $\mathcal M_\omega$. It follows that
 $\sigma_t$ is the modular group of $\varphi$ \cite{Takesaki2}.

Moreover, for each $\theta$, let $U_t(\theta)=\pi_\omega(u_t(\theta))U_t$, then
$$
U_t(\theta)\pi_\omega(a)\xi_\omega=\pi_\omega(u_t(\theta)\alpha_t(a))\xi_\omega.
$$
By continuity of the functions $F_{g,h,\theta}$,
the map $t\mapsto U_t(\theta)$ is $\sigma$-strongly continuous.
It follows that $\pi_\omega(u_t(\theta))$
is a $\sigma$- strongly continuous family of unitaries, satisfying the cocycle
condition. By Theorem 3.8 of \cite{Takesaki2},
 there are faithful semifinite normal weights $\varphi_\theta$, such that
$\pi_\omega(u_t(\theta))=[D\varphi_\theta,D\varphi]_t$.
By properties of the cocycle derivatives,
$$
\varphi_\theta(\mathbf{1})=F_{e,e,\theta}(0)=1.
$$
It follows that $\iE=(\mathcal M_\omega,\varphi_\theta: \theta\in\Theta)$ is an experiment
in $\iE(\Theta)$ and  $\omega=\omega_\iE$.

\qed

\subsection{The convex structure of experiments}

A convex combination of experiments can be obtained as follows.
Let $\iE_i=(\iA_i, \varphi_{i,\theta} :  \theta\in\Theta)$, $i=1,2$
be two experiments in $\iE(\Theta)$ and let $0<\lambda<1$.
Then we define an experiment $\iE_\lambda\in\iE(\Theta)$ by
$$
\iE_\lambda=(\iA_1\oplus\iA_2, \varphi_\theta=\lambda\varphi_{1,\theta}\oplus (1-\lambda)\varphi_{2,\theta} :\ \theta\in\Theta).
$$
It is easy to see that
$$
[D\varphi_\theta,D\varphi]_t=[D\varphi_{1,\theta},D\varphi_1]_t\oplus [D\varphi_{2,\theta},D\varphi_2]_t,\quad \theta\in\Theta,\ t\in\mathbb R
$$
and this implies that $\omega_{\iE_\lambda}=\lambda\omega_{\iE_1}+(1-\lambda)
\omega_{\iE_2}$.
We will characterize the extremal points in $\iE(\Theta)$.

\begin{theorem}\label{thm:conv}
Let $\tilde\omega$, $\omega$ be two canonical states, such that
$\tilde\omega\le t\omega$ for some $t>0$. Then there is a positive element
$T$ in the center of $\pi_\omega(G)''$, with $\|T\|\le t$, satisfying
\begin{equation}\label{eq:norm}
\omega_\theta(T)=\omega(T)=1,\qquad \forall \theta,
\end{equation}
and such that
\begin{equation}\label{eq:conv}
\tilde\omega(g)=\<\xi_\omega,\pi_\omega(g)T\xi_\omega\>,\quad g\in G
\end{equation}
Conversely, let $T\ge0$ be a central element in $\pi(G)''$ satisfying
(\ref{eq:norm}), then
(\ref{eq:conv}) defines an experiment in $\mathcal E(\Theta)$.
\end{theorem}

\noindent{\it Proof.} Let $\tilde\omega\le t\omega$, then by standard arguments there is a positive  element $T\in \pi_\omega(G)'$, $\|T\|\le t$, such that
(\ref{eq:conv}) holds.  Therefore, $\tilde\omega$ can be extended to a normal
state on $\pi_\omega(G)''$, which we again denote by $\tilde\omega$.
Let $a,b$ be elements in $\mathbb C[G]$, then by (\ref{eq:KMS}),
$$
\tilde\omega(\pi_\omega(a)\sigma_s^\omega(\pi_\omega(b)))=\<\xi_\omega,\pi_\omega(a\alpha_s(b))T\xi_\omega\>=
\tilde \omega(a\alpha_s(b)).
$$
Since $\pi_\omega(\mathbb C[G])$ is $\sigma$-strongly dense in
$\pi_\omega(G)''$,
we obtain from  Theorem \ref{thm:canon} that  $\tilde\omega$ satisfies the
modular  condition for $\sigma_t^\omega$.
This implies that  there is a positive central element $S$ in $\pi_\omega(G)''$, such that
$\tilde \omega(a)=\<\xi_\omega, aS\xi_\omega\>$ for all $a\in \pi_\omega(G)''$. Since $\xi_\omega$ is
separating for $\pi_\omega(G)'$, we have $T=S$.

To obtain the  condition (\ref{eq:norm}), let $F,\tilde F\in \mathbb A(J)$ be such
that
\begin{eqnarray*}
F(t+i)&=& \omega(T[D\omega_\theta,D\omega]_t),
\quad F(t)=
\omega_\theta([D\omega_\theta,D\omega]_tT),\\
\tilde F(t+i)&=& \tilde\omega(u_t(\theta)),\quad \tilde F(t)=\tilde \omega_{\theta}(u_t(\theta)),
\end{eqnarray*}
where we have used (\ref{eq:theta}) and the properties of the cocycle derivatives. Then
$F(t+i)=\tilde F(t+i)$ for all $t$ and this implies $F=\tilde F$. In particular,
$\omega_\theta(T)=F(0)=\tilde F(0)=1$.

Conversely, let $T\ge 0$ be a central element, satisfying (\ref{eq:norm}), then it is not difficult to check that $\tilde \omega$ given by (\ref{eq:conv}) satisfies the
properties in Theorem \ref{thm:canon}.

\qed

\begin{corollary} A canonical state  $\omega$ is extremal if and only if
the center of $\pi_\omega(G)''$ contains no positive element $T$, satisfying
$\omega_\theta(T)=\omega(T)$ for all $\theta$, other than a multiple of
identity.
\end{corollary}
{\it Proof.} Let the experiment $\omega$ be such that $\pi_\omega(G)''$ has the required property and let $\omega=\lambda\omega_1+(1-\lambda)\omega_2$.
Then $\omega_1\le \frac 1{\lambda}\omega$
and by the previous Theorem, $\omega_1$ is of the form (\ref{eq:conv}) for some positive central element $T$, satisfying (\ref{eq:norm}). It follows that
$T=I$ and  we must have $\omega_1=\omega_2=\omega$.

Conversely, suppose that there is a positive  element $\tilde T$, other that a
multiple of identity, satisfying  $\omega_\theta(\tilde T)=\omega(\tilde T)$
for all $\theta$. Then  by putting
$T=1/\omega(\tilde T)\tilde T$ in Theorem \ref{thm:conv},
we obtain an experiment $\omega_1\le t\omega$, with  $t=\|T\|>1$.
Since the vector $\xi_\omega$ is separating for $\pi_\omega(G)''$, we must have $\omega_1\ne\omega$.

It follows that  $\omega=\frac 1t\omega_1+(1-\frac1t)\omega_2$, where
$\omega_2$ has the form (\ref{eq:conv}) with the element
$S=1/(t-1)(t-T)$.  Since $S$ is a positive central element,
satisfying (\ref{eq:norm}), $\omega_2$ is an experiment.

\qed

\begin{corollary}
If $\mathcal{E}\subset\mathcal{E}(\Theta)$ is extremal then the center of
$\pi_{\omega}(G)^{\prime\prime}$ is of the form $\mathbb{C}^{d}$ with
$1\leq d\leq|\Theta |$.
\end{corollary}
\subsection{Weak and strong convergence of quantum experiments}
\label{sec.weak.strong}

The strong convergence of quantum experiments is a natural extension of
the classical convergence with respect to the Le Cam distance.

\begin{definition}\label{def.strong.convergence}
Let $\mathcal{E}:=(\mathcal A,\varphi_\theta: \theta\in \Theta)$ and
$\mathcal{F}:=(\mathcal B,\sigma_\theta: \theta\in \Theta)$ be two quantum statistical experiments. The deficiency $\delta(\mathcal{E}, \mathcal{F})$ is defined as
$$
\delta(\mathcal{E}, \mathcal{F}) = \inf_{T} \, \sup_{\theta}
\| \varphi_{\theta} \circ T - \sigma_{\theta}\| ,
$$
where the infimum is taken over all channels $T: \mathcal{B}\to \mathcal{A}$.
The Le Cam distance between $\mathcal{E}$ and $\mathcal{F}$ is
$$
\Delta(\mathcal{E}, \mathcal{F}) := \max \left( \delta(\mathcal{E}, \mathcal{F}),\ \delta(\mathcal{F}, \mathcal{E}) \right).
$$

We say that a net $\iE_\alpha:=(\mathcal A_\alpha,\varphi_{\theta,\alpha} : \theta\in\Theta)$,
$\alpha\in \mathcal{I}$, converges strongly to $\iE$ if
$\Delta(\mathcal{E}_{\alpha} , \mathcal{E}) \to 0$, i.e. there are
channels
$
T_\alpha:\mathcal A_\alpha\to\mathcal A
$
and
$
S_\alpha:\mathcal A\to\mathcal A_\alpha,
$
such that
\begin{eqnarray}
&\sup_{\theta\in\Theta}&
\|\varphi_\theta\circ T_\alpha-\varphi_{\theta,\alpha}\|\to 0\label{eq:strong1},\\
&\sup_{\theta\in\Theta}&
\|\varphi_{\theta,\alpha}\circ S_\alpha-\varphi_{\theta}\|\to 0\label{eq:strong2}.
\end{eqnarray}
We say that $\mathcal E_\alpha$ converges weakly to $\mathcal E$ if the canonical states converge pointwise
$$
\omega_{\iE_\alpha}(g)\to \omega_\iE(g),\qquad \forall g\in G.
$$
\end{definition}
\begin{theorem}\label{compactness.channels}
Let $\mathcal{A},\mathcal{B}$ be C$^{*}$-algebras and let $CP_{1}(\mathcal{B}, \mathcal{A})$ be the space of unital completely positive maps $T:\mathcal{B}\to \mathcal{A}$. Then $CP_{1}(\mathcal{B}, \mathcal{A})$ is compact with respect to
the topology defined by convergence of the linear functionals $ T\mapsto \phi( T(b))$ for all $b\in\mathcal{B}$ and $\phi \in \mathcal{A}^{*}$.
\end{theorem}

\noindent{\it Proof.} Standard application of Tychonoff's Theorem.

\qed

We will now show that the Le Cam distance is a metric on the space of equivalence classes of quantum statistical experiments.
\begin{lemma}\label{lemma:strong.equiv}
The experiments
$\mathcal{E}:= (\mathcal A,\varphi_\theta: \theta\in \Theta)$ and
$\mathcal{F}:=(\mathcal B,\sigma_\theta: \theta\in \Theta)$ are statistically equivalent
if and only if $\Delta(\mathcal{E}, \mathcal{F}) =0$.
\end{lemma}

\noindent{\it Proof.} The direct implication follows from the definitions.
We have to prove that if $\Delta(\mathcal{E}, \mathcal{F}) =0$ then there exists
a channel $T: \mathcal{B}\to \mathcal{A}$ such that
$\varphi_{\theta} \circ T = \sigma_{\theta}$ for all $\theta$, and similarly in the opposite direction.  Let $T_{\alpha}$ be a sequence (net) of channels such that
$$
\sup_{\theta} \| \varphi_{\theta} \circ T_{\alpha} - \sigma_{\theta} \| \to 0 .
$$
By Theorem \ref{compactness.channels} applied to $\mathcal{A}, \mathcal{B}$ seen as
$C^{*}$-algebras we have that $CP_{1}(\mathcal{B}, \mathcal{A})$ is compact and thus there exists a subnet $T_{ I(\alpha) }$ which converges to some  unital completely positive map $\tilde{T}$. The two statements together imply that
$\varphi_{\theta}\circ\tilde{T} = \sigma_{\theta}$.
The only problem is now that $\tilde{T}$ is not necessarily normal.

Let us denote by $\tilde{T}_*:\mathcal A_*\to\mathcal B^*$ the restriction to
$\mathcal A_*$ of the adjoint map $\tilde{T}^*$, then the map $T':=(\tilde{T}_*)^*: \tilde{\mathcal B}\to \mathcal A$
 is an extension of $\tilde{T}$ to the universal enveloping von Neumann algebra $\tilde{\mathcal B}\simeq
\mathcal B^{**}$ of $\mathcal B$. Clearly, $T'$ is completely positive and
unital. Let $z_0$ be the central projection in
$\tilde B$, such that $\mathcal B_*=\mathcal B^*z_0$, see \cite{Ta.} and let
$\psi$ be any state in $\mathcal B_*$. Define the map
$S: \mathcal B\to \tilde{\mathcal B}$ by
$$S(a)=az_0+\psi(a)(1-z_0)$$
Then $S$ is completely positive and unital, moreover,
$\varphi \circ S\in \mathcal B_*$ for all $\varphi\in\mathcal B^*$ and
$\varphi\circ S =\varphi$ for $\varphi\in \mathcal B_*$.

Finally, let $T=S\circ T'$, then $T: \mathcal B\to\mathcal A$
is a channel, such that
$\varphi_{\theta}\circ T = \sigma_{\theta}$.

\qed

We will now show that our definition of weak convergence coincides with the classical one in the case
of commutative statistical
 experiments with faithful states.
\begin{lemma}
Let $\mathcal{E}= (\Omega, \Sigma, P_{\theta} :\theta\in\Theta)$ and
$\mathcal{E}_{n}=(\Omega^{(n)}, \Sigma^{(n)},P^{(n)}_{\theta} :\theta\in\Theta)$ be classical
statistical experiments with
$n=1,2,..$. Assuming that all experiments belong to the class
$\mathcal{E}(\Theta)$, that is $P_{\theta}\sim P_{\theta_{0}}$ and
$P^{(n)}_{\theta}\sim P^{(n)}_{\theta_{0}}$ for some $\theta_{0}$, then the following
are equivalent
\begin{enumerate}
\item[(i)] $\mathcal{E}_{n}$ converges weakly to $\mathcal{E}$ in the sense of Definition \ref{def.weak.convergence.classical} for classical experiments.
\item[(ii)]  $\mathcal{E}_{n}$ converges weakly to $\mathcal{E}$ in the sense of
Definition \ref{def.strong.convergence} for quantum experiments.
\end{enumerate}
\end{lemma}

\noindent{\it Proof.}
Without loss of generality we can consider $\Theta$ to be finite.
According to Theorem \ref{th.weak.convergence} $\mathcal{E}_{n}$ converges weakly
to $\mathcal{E}$ if and only if the corresponding sequence of  likelihood ratio processes
$\Lambda^{(n)}_{\theta_{0}}$ converges in distribution to  $\Lambda^{(n)}_{\theta_{0}}$. We will show that the latter is equivalent to Definition \ref{def.strong.convergence}.
Thus we can represent all experiments as families of distributions on $\mathbb{R}_{+}^{|\Theta|-1}$  with $Q_{\theta} (d r)= r_{\theta}\lambda(d r)$ and
$Q^{(n)}_{\theta}(dr) = r_{\theta}\lambda^{(n)}(dr)$ where
$\lambda, \lambda_{n}$ are the laws of their respective likelihood ratio processes.
The associated von Neumann algebras are $\mathcal{A}:=L^{\infty}(\mathbb{R}_{+}^{|\Theta|-1},\, \lambda)$ and
$A_{n}:=L^{\infty}(\mathbb{R}_{+}^{|\Theta|-1},\, \lambda^{(n)})$ and
the cocycle derivatives act by multiplication with the function $r_{\theta}^{it}$ (for $\theta\neq \theta_{0}$):
$$
[DQ_{\theta}, D \lambda ]_{t} :f(r) \mapsto   f (r)\cdot r_{\theta}^{it}.
$$
Since by assumption, all measures have support in the interior of
$\mathbb{R}_{+}^{|\Theta|-1} $, we can consider their restriction to this subset without altering the weak convergence property (cf. Theorem 1.3.10 \cite{vanderVaart&Wellner}).
Assuming $(i)$ and considering that the functions
$$
r \mapsto \prod_{\theta\neq \theta_{0}} r_{\theta}^{it_{\theta}} ,
$$
are bounded and continuous on the interior of $\mathbb{R}_{+}^{|\Theta|-1}$, we obtain
$$
\int  \prod_{\theta\neq \theta_{0}} r_{\theta}^{it_{\theta}}   \lambda^{(n)}(dr) \to
\int  \prod_{\theta\neq \theta_{0}} r_{\theta}^{it_{\theta}}   \lambda(dr) ,\quad {\rm ~as~} n\to \infty,
$$
which proves $(ii)$.

Conversely, if $(ii)$ holds, we can map $r$ one-to-one into
$x\in\mathbb{R}^{|\Theta|-1}$ by $x_{\theta} =  \log r_{\theta}$. Then
$$
\int  \prod_{\theta \neq \theta_{0}} r_{\theta}^{it_{\theta}}   \lambda(dr) =
\int  \prod_{\theta\neq \theta_{0}} e^{i x_{\theta}t_{\theta}} \mu(dx) ,
$$
with $\mu(A) = \lambda(\log^{-1}(A))$. The right hand side represents
the characteristic function of the measure $\mu$ and by L\'{e}vy-Cram\'{e}r continuity Theorem we get that $\mu_{n}$ converges weakly to $\mu$. Finally, by the continuity of the $x\to r$ transformation we get $(i)$.

\qed

\begin{proposition}\label{prop:theta} Let $\iE_\alpha$, $\alpha\in \mathcal I$ be a net
of experiments  in $\iE(\Theta)$, converging weakly to
$\mathcal E\in\mathcal E(\Theta)$.
Let $\omega_\theta$, $\omega_{\theta,\alpha}$ be defined by
(\ref{eq:theta}). Then
$\lim_\alpha \omega_{\theta,\alpha}(g)=\omega_\theta (g)$ for all $g\in G$,
$\theta\in\Theta$.
\end{proposition}

\noindent{\it Proof.}
First, note that on the set of states, the  pointwise convergence
coincides with the weak* convergence in $L_\infty(G)$, and since the unit ball
is compact in this topology, it is enough to prove that any convergent
subnet $\omega_{\theta,\gamma}$ must converge to $\omega_\theta$.

Let $g\in G$ and let $F_\alpha:=F_{\mathcal E_\alpha,g,\theta}$,
$F:=F_{\mathcal E,g,\theta}$. Then $F_\alpha,F\in \mathbb A(J)$ and
$|F_\alpha(z)|\le 1$, $|F(z)|\le 1$ for $z\in J$. By assumptions,
$F_\alpha(t)$ converges to $F(t)$ for each $t$.
We will use the following family of functions:
$$
f_{\beta,z}(t)=\frac 1{\sqrt{\beta\pi}}\exp\{-\frac {(t-z)^2}{\beta}\},\qquad
\beta>0,\ z\in\mathbb C
$$
For any $\beta>0$, define
$$
\phi_{\alpha,\beta}(z)=\int F_\alpha(t)f_{\beta,z}(t)dt,\qquad  \phi_{\beta}(z)=\int
F(t)f_{\beta,z}(t)dt
$$
Then $\phi_{\alpha,\beta}$, $\phi_\beta$ are entire analytic and uniformly bounded
on compact subsets in $\mathbb C$. Moreover, for $s\in \mathbb R$,
$$
|\phi_{\alpha,\beta}(s)-\phi_\beta(s)|\le \int |F_\alpha(t)-F(t)|f_{\beta,s}(t)
dt\to 0
$$
by dominated convergence theorem. It follows that
$\phi_{\alpha,\beta}(z)\to\phi_\beta(z)$ for all $z\in\mathbb C$.

Since $F_\alpha$, $F$ are analytic in $J$ and continuous on $\bar J$,
$$
\phi_{\alpha,\beta}(z+w)=\int F_\alpha(t+w)f_{\beta,z}(t)dt,\quad
\phi_{\beta}(z+w)=\int F(t+w)f_{\beta,z}(t)dt,\qquad w\in \bar J
$$
In particular,
$$
\int F_\alpha(t+i) f_{\beta,0}(t)dt=\phi_{\alpha,\beta}(i)\to
\phi_\beta(i)=\int F(t+i)f_{\beta,0}(t)dt,\qquad \beta>0
$$

Suppose now that $\omega_{\theta,\gamma}$ is a convergent subnet, then
$F_{\gamma}(t+i)=\omega_{\theta,\gamma}(u_t(\theta)g)$ converges pointwise
to some function $\psi(t)$, bounded by 1. But then again, we have
$\int F_{\gamma}(t+i)f_{\beta,0}(t)dt\to\int \psi(t)f_{\beta,0}(t)dt$,
so that  $\int \psi(t)f_{\beta,0}(t)dt=\int F(t+i)f_{\beta,0}(t)dt$ for all
$\beta>0$. Letting $\beta\to 0$, we get
$\lim_\gamma\omega_{\theta,\gamma}(g)\to \omega_\theta(g)$.

\qed

\noindent {\bf Remark.} Let us choose another point $\theta\in\Theta$ instead
of
$\theta_0$ in the definition of the canonical state. Then by the chain rule
for the cocycle derivatives,
$$
[D\varphi_{\theta'},D\varphi_\theta]_t=[D\varphi_{\theta'},D\varphi]_t
[D\varphi,D\varphi_\theta]_t,\quad \theta'\in\Theta,\ t\in\mathbb R
$$
so that we obtain the same group $G(\Theta)$ and the new canonical state is
equal to $\omega_\theta$. The above proposition implies that weak convergence
of experiments does not depend from the choice of $\theta_0$.

 We have shown that our  definition of weak convergence
corresponds to the classical one, in commutative case. What is still missing is the relation to the strong convergence, namely that weak and strong convergence are equivalent for finite parameter sets
 (cf. Theorem \ref{th.strong.equivalent.weak.finite}). Note that this would
 also  imply that strong convergence is  stronger than the weak one. We will show this equivalence under some conditions. First, we will consider
uniformly dominated sets of experiments.

Let $\psi$ be any experiment in $\mathcal E(\Theta)$ and let $B>0$. Let
us denote by $\mathcal E(\psi,B)$ the set of all experiments
$\omega\in\mathcal E(\Theta)$, such that $\omega\le B\psi$.
 By  Theorem \ref{thm:conv}, there is a one-to-one correspondence
between $\mathcal E(\psi,B)$ and the set $\mathcal Z(\psi,B)$ of
positive elements in the center of $\mathcal M_\psi$  with
$\|T\|\le B$ and  $\psi_\theta(T)=1$, for all $\theta$. Namely, for any 
$\omega\in\mathcal E(\psi,B)$, there is an
element $T\in\mathcal Z(\psi,B)$, such that
$$
\omega_\theta(g)=\psi_\theta(Tg)=\<\xi_{\psi_\theta},T\pi_\psi(g)\xi_{\psi_\theta}\>,\qquad g\in G, \theta\in\Theta
$$
and since $\psi$ is  faithful on $\mathcal M_\psi$, such $T$ is unique.
This also implies
that $\omega_\theta$ can be extended to a normal state  on $\mathcal M_\psi$.

Let us endow $\mathcal E(\psi,B)$ with the topology of pointwise convergence and
$\mathcal Z(\psi,B)$ with the $\sigma(\mathcal M_\psi,\mathcal M_{\psi*})$- topology.
Then  $\mathcal Z(\psi,B)$ is compact.
Let $T_\alpha$ be a  net in $\mathcal Z(\psi,B)$, converging to $T$
and let  $\omega_\alpha$ and
$\omega$ be the corresponding canonical states  in $\mathcal E(\psi,B)$. Then for any $g\in G$,
$$
\omega_\alpha(g)=\<\xi_\psi, T_\alpha\pi_\psi(g)\xi_\psi\>\to
\<\xi_\psi, T\pi_\psi(g)\xi_\psi\>=\omega(g),
$$
so that the map $\Psi: \mathcal Z(\psi,B)\ni T\mapsto
\psi(T\cdot)\in\mathcal E(\psi,B)$ is continuous. It follows that $\mathcal E(\psi,B)$ is compact.

Conversely, let $\omega_\alpha$ be a net in $\mathcal E(\psi,B)$, converging  to
$\omega$ and let $T_\alpha$, $T$ be the corresponding elements in
$\mathcal Z(\psi,B)$. Then for any $a,b\in \mathbb C[G]$, we have
$$
\<\pi_\psi(a)\xi_\psi,T_\alpha \pi_\psi(b)\xi_\psi\>= \omega_\alpha(a^*b)\to
\omega(a^*b)= \<\pi_\psi(a)\xi_\psi,T \pi_\psi(b)\xi_\psi\>
$$
Since the vectors
$\pi(a)\xi_\psi$, $a\in\mathbb C[G]$ are dense in $H_\psi$ and $T_\alpha$ are
uniformly bounded, this implies that $T_\alpha$ converges to $T$. It follows that
the inverse map $\Psi^{-1}: \mathcal E(\psi,B)\to \mathcal Z(\psi,B)$ is
continuous. Moreover,  we get that 
$\omega_{\theta,\alpha}(a)\to \omega_\theta(a)$, for all $a\in\mathcal M_\psi$,
$\theta\in\Theta$.

We can summarize as follows.

\begin{lemma}\label{lemma:continuous} The topology in $\mathcal E(\psi,B)$ coincides with the topology obtained from
the weak topology in $\mathcal M_{\psi*}$. The set $\mathcal E(\psi,B)$ is compact, and therefore sequentially
compact, by the Eberlein - Smulyan theorem.
\end{lemma}
Now we can state the equivalence theorem, for uniformly dominated sequences of
experiments of type I with discrete center.
\begin{theorem}\label{th.weak&strong}
Let  $\mathcal{E}_n:= (\mathcal{A}_{n},
\varphi_{\theta,n} : \theta\in\Theta)$ be a sequence of experiments in
$\mathcal E(\Theta)$ with $\Theta$ a finite set.
Assume that the  sequence is uniformly dominated, i.e. the canonical states
$\omega_n:=\omega_{\mathcal  E_n}\in \mathcal E(\psi,B)$ for all $n$,
for some  fixed experiment  $\psi$ and  $B>0$.
Assume further that the minimal sufficient von Neumann
algebras of the experiments  $\mathcal{E}_{n}$ are type I with discrete center.
Then  $\mathcal E_n$ converges weakly to $\mathcal E$ if and only if
 $\Delta(\mathcal{E}_{n}, \mathcal{E})\to 0$, i.e. there exist sequences
of channels
$$
\alpha_n:\mathcal A_n\to \mathcal A , \quad \beta_n:\mathcal A\to \mathcal
A_n,
$$
such that
$$
\lim_{n\to\infty}\|\varphi_\theta\circ \alpha_n-\varphi_{\theta,n}\|=0,\quad
\lim_{n\to\infty}\|\varphi_{\theta,n}\circ \beta_n-\varphi_\theta\|=0 ,
\qquad \forall \theta \in \Theta.
$$
\end{theorem}

\noindent{\it Proof.} Let $\omega_n\in\mathcal E(\psi,B)$, satisfying the assumptions.
 Then the support $p_n$ of $\omega_n$ in $\mathcal M_\psi$ is a central projection, such that
 $p_n\mathcal M_\psi$ is type I with discrete center.
Let $\bar{\omega}=\sum_n\lambda_n\omega_n$, with some $\lambda_n>0$,
$\sum_n\lambda_n=1$, then  $\bar{\omega}$ is an experiment in
$\mathcal E(\psi,B)$. Let  $p$ be the support of $\bar{\omega}$, then $\mathcal M_{\bar{\omega}}\simeq
p\mathcal M_\psi$ and   $p=\sup_n p_n$.
It follows that $\mathcal M_{\bar{\omega}}$ is type I with discrete center
and $\omega_n(a)=\omega_n(pa)$ for $a\in \mathcal M_\psi$. Moreover, since 
$\omega_{\theta,n}$ have the same support for all $\theta$, 
$\omega_{\theta,n}(pa)=\omega_{\theta,n}(a)$.

Suppose that $\mathcal E_n \to \mathcal E$ weakly and let $\omega:=\omega_{\mathcal E}$. By the remarks before 
Lemma \ref{lemma:continuous}, the normal extensions of $\omega_{\theta,n}$ 
converge weakly to $\omega_\theta$ in
$\mathcal M_{\psi*}$. It follows that $\omega_\theta(pa)=\omega_\theta(a)$ for
all $a\in \mathcal M_\psi$ and 
 we can conclude that  $\omega_{\theta,n}(a)\to \omega_\theta(a)$,
for all $a\in \mathcal M_{\bar{\omega}}$ and $\theta\in\Theta$.

In \cite{DellÕAntonio} it is shown that the preduals of the type I von Neumann algebras with atomic center have
 the Kadec-Klee property: any sequence of normal states
$\omega_{n}$ converging weakly to a normal state $\omega$ is also norm convergent
$\lim_n\| \omega_{n} - \omega\|=0$. We apply this to the experiments
$\tilde{\mathcal{E}}_n :=(\mathcal M_{\bar{\omega}}, \omega_{n,\theta}:\theta \in \Theta)$ which by
construction are equivalent with the original experiments $\mathcal{E}_n$ and we
get $\lim_n\|\omega_{n,\theta} - \omega_\theta\| =0$ for all $\theta$.

Conversely, suppose that $\Delta(\mathcal E_n,\mathcal E)\to 0$. By Lemma
\ref{lemma:continuous}, there is a subsequence  $\mathcal E_{n_k}$, converging
weakly to some experiment $\mathcal F$. By the first part of the proof, 
$\Delta(\mathcal E_{n_k},\mathcal F)\to 0$. Since also 
$\Delta(\mathcal E_{n_k},\mathcal E)\to 0$, we have $\Delta(\mathcal E,\mathcal F)=0$  and by Lemma  \ref{lemma:strong.equiv},
this implies that  $\mathcal F$ is equivalent with $\mathcal E$, so that 
$\omega_{\mathcal F}=\omega$. 
It follows that the whole sequence converges weakly to $\mathcal E$.

\qed

\noindent{\bf Remark.} Our result is complementary to the classical one in two respects. First,  the range of covered experiments consists of type I algebras with discrete
center, thus the ``typical'' noncommutative probability spaces. Second, the
proof uses the Kadec-Klee property specific to this type of algebras and not true for general probability spaces.

\section{Quantum Central Limit Theorem}\label{sec.qclt}

We have seen that in classical statistics the Central Limit Theorem is an essential ingredient of the proof of local asymptotic normality in its weak version. In the quantum case the situation is similar, so we will proceed in this
section to explain the quantum Central Limit Theorem in the simplest situation, that is for a matrix algebra $M(\mathbb{C}^{d})$ and a faithful state $\varphi$ on $M(\mathbb{C}^{d})$, i.e. a state whose density matrix $\rho$
is strictly positive. However the result holds in the general framework of $C^{*}$-algebras and we refer
to the references \cite{Ohya&Petz, Petz} for more details and proofs.


Let $L^{2}(\rho) = (M(\mathbb{C}^{d}) , \langle\cdot, \cdot \rangle_{\rho})$ be the
{\it complex} Hilbert space with inner product
$$
\langle X, Y \rangle_{\rho}  = \mathrm{Tr}(\rho  Y^{*} X ), \qquad X, Y\in M(\mathbb{C}^{d}).
$$
On $M(\mathbb{C}^{d})$ we define the symplectic form
$\sigma$ by
$$
\sigma(X, Y) = \mathrm{Im}(\langle X, Y \rangle_{\rho}),
$$
and we construct the algebra $CCR(M(\mathbb{C}^{d}), \sigma)$ of canonical commutation relations having as generators the Weyl operators $W(X)$ for all
$X\in M(\mathbb{C}^{d})$ and satisfying the relations
$$
W(X)W(Y) = W(X+Y) \exp( -i \sigma(X, Y)).
$$
On this algebra we define the quasifree state $\phi$ by
$$
\phi(W(X)) = \exp(- \frac{1}{2} \alpha (X, X) )  .
$$
where $\alpha$ is the positive bilinear form
$
\alpha(X, Y) = \mathrm{Re} (\langle X, Y \rangle_{\rho}).
$
By the GNS construction, $\phi$ generates a representation of the CCR algebra and
for now we denote by $W(X)$ the Weyl operators in this representation and occasionally express them in terms of the field operators $W(X)= \exp(i B(X))$.
Note that any field operator $B(X)$ has a Gaussian distribution centered at $0$ and with variance $  \|X \|_{\rho}^{2}=\alpha(X,X)$.

Consider the tensor product $\bigotimes_{k=1}^{n}M(\mathbb{C}^{d})$ of algebras $M(\mathbb{C}^{d})$ which is generated by elements of the form
\begin{equation}\label{eq.xk}
X^{(k)} = \mathbf{1}\otimes \dots \otimes X \otimes \dots \otimes \mathbf{1},
\end{equation}
with $X$ acting on the $k$-th position of the tensor product. We are interested in the asymptotics as $n\to\infty$ of the joint distribution under the state $\varphi^{\otimes n}$, of `fluctuation' elements of the form
$$
F_{n}(X) :=\frac{1}{\sqrt{n}} \sum_{k=1}^{n} X^{(k)}.
$$
\begin{theorem}\label{th.clt}
Let $A_{1}, \dots , A_{s}\in M(\mathbb{C}^{d})^{sa}$ satisfying $\varphi(A_{l})=0$, for $l=1,\dots , s$. Then we have the following
\begin{eqnarray*}
&&
\lim_{n\to\infty} \varphi^{\otimes n} \left(\prod_{l=1}^{s} F_{n}(A_{l}) \right) =
\phi \left( \prod_{l=1}^{s}\left( B(A_{l}) \right)\right),\\
&&
\lim_{n\to\infty} \varphi^{\otimes n} \left( \prod_{l=1}^{s} \exp( iF_{n}(A_{l}) ) \right) =
\phi\left( \prod_{l=1}^{s} W( A_{l} )  \right).
\end{eqnarray*}
\end{theorem}
Note that only joint distributions for selfadjoint operators are considered. This is
sufficient for the purpose of this paper and for the rest of this section we concentrate on the properties of the subalgebra $CCR(M(\mathbb{C}^{d})^{sa}, \sigma)$ generated by the Weyl operators  $W(A)$ with $A$ selfadjoint operator in $M(\mathbb{C}^{d})$. This subalgebra will be the key to understanding the limit quantum experiment.

In the case of selfadjoint operators the symplectic form becomes
$$
\sigma(A, B) = \frac{i}{2} \mathrm{Tr} \left( \rho [A, B] \right).
$$
The bilinear form $\alpha$ is a positive inner product on
$M(\mathbb{C}^{d})^{sa}$ and from now on we will denote its restriction to this
subspace as
$$
(A,B)_{\rho} := \alpha(A, B)  =
\mathrm{Tr} \left( \rho A\circ B\right),
$$
and the corresponding real Hilbert space by $L^{2}_{\mathbb{R}} (\rho) =(M(\mathbb{C}^{d})^{sa}, (\cdot, \cdot )_{\rho})$. We write $L^{2}_{\mathbb{R}}(\rho)$ as a direct sum of orthogonal subspaces
$\mathcal{H}_{\rho} \oplus \mathcal{H}_{\rho}^{\perp}$ where
$$
\mathcal{H}_{\rho}= \left\{ A \in L^{2}_{\mathbb{R}}(\rho)   : [A, \rho ] =0 \right\}.
$$
In particular if $B= B_{1} \oplus B_{2}\in L^{2}_{\mathbb{R}}(\rho)$ then
\begin{equation}\label{eq.state.factorization}
\phi(W(B)) =  \exp\left(   -\frac{1}{2} (B_{1}, B_{1})_{\rho} \right) \exp\left(   -\frac{1}{2} (B_{2}, B_{2})_{\rho} \right) .
\end{equation}
Moreover since $\sigma(A, B)=0$ for $A\in \mathcal{H}_{\rho}$ and $B$ arbitrary
we get the following factorization
\begin{equation}\label{eq.factorization.1}
CCR( M(\mathbb{C}^{d})^{sa}, \sigma ) \cong
CCR(  \mathcal{H}_{\rho}, \sigma ) \otimes CCR( \mathcal{H}_{\rho}^{\perp}, \sigma ),
\end{equation}
and by (\ref{eq.state.factorization}) the state $\phi$ factorizes as
\begin{equation}\label{eq.factorization.2}
\phi= \phi_{1} \otimes \phi_{2}.
\end{equation}
The left side of the tensor product is a commutative algebra which is isomorphic to $L^{\infty}\left(\mathbb{R}^{|\mathcal{H}_{\rho}|}\right)$ carrying a Gaussian state with
covariance $(A, B)_{\rho}$.

\section{Local asymptotic normality for quantum states}\label{sec.qlan}
We are now ready to introduce the central result of the paper which extends the concept of local asymptotic normality to the quantum domain and provides also an important example of convergence of quantum statistical experiments. Throughout this section
we consider the algebra $\mathcal A=M_d(\mathbb C)$, a family of strictly positive density matrices $\rho_{\theta}$ in $M_d(\mathbb C)$ such that the map
$\theta\mapsto \rho_{\theta}$ has the property that both the eigenvalues and eigenvectors of $\rho_{\theta}$ are twice continuously differentiable, and denote by
$\varphi_\theta$ the corresponding faithful  states on $\mathcal A$.

Consider $n$ quantum systems prepared in the same state
$\varphi_{\theta}$ with $\theta\in \Theta\subset\mathbb{R}^{m}$ an unknown parameter which will be taken of the form $\theta= \theta_{0}+ u/\sqrt{n}$ where $u$ is an
unknown parameter belonging to some open, bounded neighborhood of the origin $I\subset \mathbb R^m$, and $\theta_{0}$ is a fixed and known parameter.
We are interested in the asymptotic behavior as $n\to\infty$ of the quantum
statistical experiments
$$
\mathcal{E}_{n} = \left( \mathcal A_n=\mathcal A^{\otimes n}, \varphi_{u,n}=
(\varphi_{\theta_{0}+ u/\sqrt{n}} )^{\otimes n} : u\in I \right) ,
$$
whose family of states is indexed by a parameter $u\in I$.
Namely, we will show that the sequence $\iE_n$ converges weakly to an experiment $\iE$, consisting of a family $\{\phi^u,u\in I\}$
of quasifree states on the CCR algebra $\left(M(\mathbb{C}^{d})^{sa},\sigma
\right)$ with $\sigma(A,B) = \frac{i}{2}\mathrm{Tr}(\rho_{\theta_{0}} [A,B])$ (cf. Section \ref{sec.qclt}).

\subsection{One parameter unitary family of states}\label{sec.unitary.one.parameter.lan}
We will first consider a simple model of a one-parameter family of states where the eigenvalues of the density matrices are fixed and only the eigenvectors vary smoothly.
This will be helpful in the next section where the general multi-parameter case is considered and it is shown that the quantum local asymptotic normality can be obtained by combining the fixed eigenvalues situation with the classical problem of evaluating the eigenvalues of a density matrix for fixed eigenvectors.


For simplicity we consider a local neighborhood around $\theta_{0}=0$.
Let $\rho=\rho_{\theta_{0}}$ be a density matrix on $\mathcal{A}:=M(\mathbb{C}^{d})$ and define $\rho_{a} = e^{iaH} \rho e^{-iaH}$  for $a\in \mathbb{R}$, where $H$ is a selfadjoint operator which can be chosen such that $\varphi (H) =0$.
Denote by $\varphi_{a}$ the corresponding state functionals
$\varphi_{a}(A) := \mathrm{Tr}(\rho_{a} A)$. Consider now $n$ quantum systems prepared in the same state $\rho_{u/\sqrt{n}}$ where $u$ is an unknown parameter belonging to some bounded open
interval $I\subset\mathbb{R}$ containing the origin. We are interested in the asymptotic behavior as $n\to\infty$ of the quantum statistical experiments
\begin{equation}\label{eq.en}
\mathcal{E}_{n} = \left( (M(\mathbb{C}^{d}))^{\otimes n}, \rho_{u/\sqrt{n}}^{\otimes n} : u\in I \right) ,
\end{equation}
whose family of states is indexed by a parameter $u\in I$.

As explained in Section \ref{sec.lan}, the likelihood ratio process is a sufficient statistic in the case of classical statistical experiments, and the local asymptotic normality property means that this process converges in distribution to the corresponding likelihood process of the limit experiment. For a quantum experiment however, there is no obvious
analogue of the likelihood ratio process. In Section \ref{sec.quantum.sufficiency} we argued that the guiding principle in finding the quantum analog of this process should be to look at operators which are intrinsically related to the quantum experiment in the sense that they generate the minimal sufficient algebra, similarly to the case of the likelihood ratio process. Such operators are the Connes cocycles which in the case of the experiment $\mathcal{E}_{n}$ are given by
$$
C^{(n)}_{u,t}=
[D \varphi_{{u/\sqrt{n}}}^{\otimes n} ,\, D \varphi^{\otimes n} ]_{t}:=
\left[ \rho_{u/\sqrt{n}}^{\otimes n}\right]^{it}
\left[ \rho^{\otimes n}\right]^{-it}.
$$
We can rewrite this as
\begin{eqnarray*}
C^{(n)}_{u,t}=
&&
\left\{\left[e^{i u H /\sqrt{n}}  \rho  e^{-i u H /\sqrt{n}} \right]^{\otimes n}  \right\}^{it}
\left[ \rho^{\otimes n}\right]^{-it}=\\
&&
\left[e^{i u H /\sqrt{n}}   \rho^{it}  e^{-i u H /\sqrt{n}} \right]^{\otimes n}
\left[ \rho^{-it} \right]^{\otimes n}=\\
&&
\left[e^{i u H /\sqrt{n}}  \rho^{it}  e^{-i u H /\sqrt{n}} \rho^{-it} \right]^{\otimes n}  =\\
&&
\left[e^{i u H /\sqrt{n}}  e^{-i u \sigma_{t}(H) /\sqrt{n}}  \right]^{\otimes n}=\\
&&
\exp\left(\frac{i u}{\sqrt{n}} \sum_{k=1}^{n} H^{(k)} \right)
\exp\left( \frac{-i u}{\sqrt{n}} \sum_{p=1}^{n}\sigma_{t}( H)^{(p)}  \right).
\end{eqnarray*}
where
$\sigma_{t}(H):=
 \rho^{it}
H
\rho^{-it}
$
is the action of the modular group of $\varphi$ on $H$, and
$H^{(k)}$ represents the operator $\mathbf{1}\otimes \dots \otimes H \otimes \dots \otimes \mathbf{1}$ with $H$ acting on the $k$-th term of the tensor product.

Consider now the expectation values of products of such cocycles with respect to the state $\varphi^{\otimes n}$:
\begin{eqnarray*}
&&
E^{(n)}(u_{1}, t_{1},\dots,  u_{s} , t_{s}) :=
 \varphi^{\tens n}
\left[  \prod_{l=1}^{s} C^{(n)}_{u_{l}, t_{l}}\right] \\
&&
=\mathrm{Tr}
\left[  \rho^{\otimes n}
\prod_{l=1}^{s}
\exp\left( iu_{l} F_{n}(H) \right)
\exp\left( -i u_{l} F_{n}(\sigma_{t_{l}}(H))  \right)
\right].
\end{eqnarray*}
We apply now the second part of the central limit Theorem \ref{th.clt} to obtain
$$
\lim_{n\to \infty} E^{(n)}(u_{1}, t_{1}, \dots ,u_{s} , t_{s})=
\phi
\left(
\prod_{l=1}^{s}
W\left( u_{l} H  \right) W \left(- u_{l} \sigma_{t_{l}}(H)
\right)
\right),
$$
where $\phi$ is the quasifree state on the algebra
$CCR(M(\mathbb{C}^{d})^{sa},\sigma)$  with symplectic form
$$
\sigma(A, B) := \frac{i}{2}\mathrm{Tr} (\rho [A, B]).
$$
The state $\phi$ is defined by
 $
 \phi \left( W(X) \right) = \exp\left(-\frac{1}{2} \alpha(X,X) \right),
 $
where $\alpha$ is the real symmetric positive bilinear form
$
\alpha(A, B)= \mathrm{Tr} \left( \rho A \circ B \right),
$
where $A\circ B= AB+BA/2$. By using the Weyl relations we get
$$
\lim_{n\to \infty} E^{(n)}(u_{1}, t_{1}\dots u_{s} , t_{s})=
\phi\left(
\prod_{l=1}^{s}
W\left( u_{l} (H- \sigma_{t} (H)) \right)
\exp\left[ \frac{u_{l}^{2}}{2} \varphi\left([ H, \sigma_{t} (H) ] \right)\right]
\right).
$$

In analogy to the classical local asymptotic normality, we would like to interpret the expression on the right side as the expectation of a product of cocycles of the form $[D \phi^{u} , D\phi^{0} ]_{t}$ for some family of states
$\left\{ \phi^{u} : u\in I \right\}$ with $\phi^{0} =\phi$, on $\mathcal W:=CCR(M(\mathbb{C}^{d})^{sa},\sigma)$. Later on we will restrict our attention to the minimal sufficient subalgebra which is generated by the Connes cocycles \cite{Petz&Jencova} and still have a statistically equivalent quantum experiment. Let us define the family of translated states on $\mathcal W$
$$
\phi^{u}(W(A)) = \phi\left( W(u H ) W(A) W(- u H ) \right), \qquad A\in M(\mathbb{C}^{d})^{sa}.
$$
The cocycles can be calculated (see e.g. page 160 of \cite{Ohya&Petz}):
\begin{equation}\label{eq.cocycle.unitary}
[D \phi^{u} , D \phi^{0}]_{t} = W(u(H-  \sigma_{t} (H))) \exp\left[ \frac{u^{2}}{2} \varphi
\left([ H, \sigma_{t} (H) ] \right)\right].
\end{equation}
Thus we obtain the convergence in distribution of the Connes cocycles
$$
\lim_{n\to \infty}
\varphi^{\tens n}
\left(  \prod_{l=1}^{s} \left[
D \varphi_{u_{l}/\sqrt{n}}^{\tens n} \, ,\,
D \varphi^{\tens n}\right
]_{t_{l}} \right)
=
 \phi\left(
\prod_{l=1}^{s} \left[D \phi^{u_{l}} , D \phi^{0} \right]_{t_{l}} \right).
$$
Notice that $[D \phi^{u} , D \phi^{0}]_{t}$ do not commute for different times as in
general
$
\varphi \left( [H-\sigma_{t}(A) , H-\sigma_{s}(H) ]\right) \neq 0.
$ This implies that the minimal sufficient algebra $\mathcal{W}_{0}\subset \mathcal{W}$ is non-commutative and is generated by the Weyl operators $W(A)$ with
$A \in K := \mathrm{Lin}_{\mathbb{R}}( H -\sigma_{t} (H) : t\in \mathbb{R})$.
We denote by $\mathcal{E}$ the limit experiment in its minimal form
\begin{equation}\label{eq.limite}
\mathcal{E} = \left(  \mathcal{W}_{0} , \phi^{u} : u\in I \right).
\end{equation}
\begin{theorem}
As $n\to\infty$ we have
$$
\mathcal{E}_{n} \to \mathcal{E},
$$
in the sense of weak convergence of experiments,
where $\mathcal E_{n}$ is the sequence defined in \eqref{eq.en} and $\mathcal{E}$ is the
quantum Gaussian shift experiment defined in \eqref{eq.limite}.
\end{theorem}
We will take now a closer look at the limit experiment and in particular at the optimal measurement for estimating the unknown parameter $u\in I$.
It is known \cite{Holevo} that asymptotically the optimal procedure for
$\mathcal{E}_{n}$ is to measure the symmetric logarithmic derivative $\mathcal{L}$ at the point $\theta_{0}=0$ on each of the individual systems separately.
As we will see, the optimal procedure for the limit experiment is
to measure the corresponding observable $B(\mathcal{L})$ and
obtain a classical experiment with Fisher information equal to the quantum Fisher information of $\mathcal{E}$ (see also \cite{Hayashi&Matsumoto}).

 Let $A$ be an arbitrary element of $K$. When restricted to the commutative algebra generated by the field $B(A)$, the states $\phi^{u}$ give rise to a family of displaced Gaussian distributions on $\mathbb{R}$
$$
P_{A}^{u} := N(  - i u \varphi( [H, A]), \, \varphi (A^{2})).
$$
Indeed the expected value of $B(A)$ is
\begin{eqnarray*}
\phi^{u} (B(A))
&=&
\phi ( W(u H) B (A) W( -uH) ) =
\phi( B(A) + 2u \sigma(H, A) \mathbf{1}) \\
&=&
 - i u \varphi ( [H, A]),
\end{eqnarray*}
and the variance is
$
\phi(B(A)^{2}) = \alpha(A, A) = \varphi (A^{2}).
$
It can be shown that for a Gaussian shift family $( N(au , v), u\in I)$ the Fisher information is given by
$I= a^{2}/v^{2}$, thus in our case we have
\begin{equation}\label{eq.info.quadrature.A}
I_{A} = \varphi ([H, A])^{2} /\varphi (A^{2}).
\end{equation}
Coming back to the original quantum experiment $(M(\mathbb{C}^{d}), \varphi_{a} : a\in \mathbb{R})$  we define the symmetric logarithmic derivative at $\theta_{0}=0$ by
\begin{equation}\label{eq.sld}
\mathcal{L} \circ \rho =  \left.\frac{d\rho_{a} }{da} \right\vert_{a=0} = i[H, \rho].
\end{equation}
Thus
$$
i \phi ([A,H]) =
i \mathrm{Tr} (\rho [A, H]) =
i \mathrm{Tr} (A  [H, \rho] ) =\mathrm{Tr} (\rho A\circ \mathcal{L} )  = \left( A,  \mathcal{L}\right)_{\rho},
$$
and by inserting into (\ref{eq.info.quadrature.A}) we get
$
I_{A} =  |\left( A , \mathcal{L} \right)_{\rho} |^{2}/ \|A\|^{2},
$
which takes its maximum value for $A = \mathcal{L}$ Thus
$$
\sup_{A} I_{A} = I_{\mathcal{L}} = \mathrm{Tr} (\rho  \mathcal{L}^{2} ) ,
$$
where the last expression is the quantum Fisher information
$H(\rho)$ \cite{Holevo}.

We will show now that $\mathcal{L}$ belongs to the subspace $K$, so that its corresponding field belongs to the minimal sufficient algebra $\mathcal{W}_{0}$. Let $\rho= \sum_{i=1}^{d}\lambda_{i}P_{i}$ be the spectral decomposition of  $\rho$, then the symmetric logarithmic derivative can be written as
\begin{equation}\label{eq.sld2}
\langle e_{i} ,\mathcal{L} e_{j} \rangle =
2i \frac{\lambda_{i} -\lambda_{j}}{\lambda_{i}+\lambda_{j}} \langle e_{i} ,H e_{j} \rangle.
\end{equation}

By derivating $H-\sigma_{t}(H)$ with respect to $t$ we obtain that the multiple commutators $C_{r}:= [\dots[ H, \log\rho ], \dots, \log \rho]$ belongs to $K$ for any number $r$ of commutators. It is easy to see that
$$
\langle e_{i} , C_{r} e_{j} \rangle =
\langle e_{i}, H e_{j} \rangle (\log (\lambda_{j} /\lambda_{i} ))^{r},
$$
and by writing (\ref{eq.sld2}) in the form
$$
\langle e_{i} , \mathcal{L}  e_{j} \rangle  = 2i
\langle e_{i}, H e_{j} \rangle
\frac{1- e^{\log (\lambda_{j}/\lambda_{i}) }}{ 1+ e^{\log(\lambda_{j} /\lambda_{i} )}},
$$
we see that $\mathcal{L}$ belongs to the linear span of $C_{r}$
for $r\geq 1$ and thus $\mathcal{L}\in K$.

In conclusion there exists a measurement on the limit experiment such that the Fisher information of the measurement results achieves the upper bound given by the quantum Fisher information. This suggests that the {\it classical} statistical experiment
$$
\mathcal{F} = ( \mathbb{R} , P_{\mathcal{L}}^{u} : u\in I),
$$
`contains all the information' about the asymptotics of the sequence
$\mathcal{E}_{n}$. We will show that this is not true in the sense that $\mathcal{F}$ is {\it not} equivalent to $\mathcal{E}$. Indeed if that was the case there would exist a linear positive map $S$ from $L^{1}(\mathbb{R}) $ to $\mathcal{W}_{0*}$, the space of normal functionals on $\mathcal{W}_{0}$ such that
$$
S: P_{\mathcal{L}}^{u} \mapsto \phi^{u},\qquad  u\in I.
$$
But $S$ is completely positive and thus $\mathcal{E}$ and $\mathcal{F}$ can be obtained from each other by quantum randomizations which is impossible as their minimal sufficient subalgebras cannot be isomorphic \cite{Petz&Jencova}. In particular this means that there exists a classical statistical decision problem for which the minimax risk of the experiment $\mathcal{E}$ is strictly smaller than the minimax risk of the experiment $\mathcal{F}$. An example of such decision problem \cite{Guta&Kahn}, is that of distinguishing between two states $\phi^{u}$ and
$\phi^{-u}$ with $u\neq 0$ for which the optimal measurement is different from the measurement of $\mathcal{L}$.

\subsection{Local asymptotic normality: general case}\label{sec.lan.general}

We pass now the the general case of an $m$ dimensional family of states as described
in the beginning of Section \ref{sec.qlan}.  The main ingredients of the proof are the quantum central limit theorem and the following form of the law of large numbers \cite{Ohya&Petz, Petz}:

Let $\mathcal B$ be the infinite tensor product of copies of $\mathcal A$ and let $\psi$ be
the product state $\psi=\varphi\otimes\varphi\otimes\dots$
Each element $a\in\mathcal A_n$ can  be identified with the element
$a\otimes I\otimes I\otimes\dots$ in $\mathcal B$. For $a\in \mathcal A$,
we denote
$$
S_n(a):=\frac 1 n\sum_{k=0}^n a^{(k)} \in \mathcal A_n.
$$
with $a^{(k)}$ as in equation \eqref{eq.xk}, and similarly for any element
$b\in \mathcal{A}_{n}$ we denote the $k$-places translated
$b^{(k)}:= \mathbf{1}\otimes \dots \otimes \mathbf{1} \otimes b \otimes \mathbf{1}\otimes\dots \in\mathcal{B}$ .

Let us consider the GNS representation of $\mathcal B$ with respect to $\psi$
on a Hilbert space $H$ with cyclic vector
 $\Psi$. We define the contraction $V:\ H\to H$ by
 $Vb\Psi=b^{(1)}\Psi$, for $b\in \mathcal A_n$.
 Then we  have
$$
\lim_n \frac 1n \sum_{k=0}^nV^{k} a\Psi=\lim_nS_n(a)\Psi=\varphi(a)\Psi,
$$
for all $a\in \mathcal A$. As a consequence, we get the following Lemma.
\begin{lemma} \label{lemma:pom}
Let $a_n,a$ be selfadjoint elements in $\mathcal A$, such that $a_n\to a$, and let  $\tilde {\rho}_n\in \mathcal A_{*}$ be density matrices such that
$\tilde{\rho}_n\to \rho$. Let $u_n, v_n\in \mathcal A$  be unitaries such that
$u_{n}\to \mathbf{1}$ and $v_n\to \mathbf{1}$.
With the notation
$$
w_{n,t}=\exp\{it(\log \tilde{\rho}_n+\frac 1n a_n)\}\tilde{\rho}_n^{-it},\qquad t\in\mathbb R,
$$
we have
$$
\lim_{n\to\infty}
\varphi^{\otimes n}(u_{n}^{\otimes n}\, w_{n,t}^{\otimes n}\, v_n^{\otimes n})=
\exp\{it\varphi(a)\}\lim_{n\to\infty}\varphi^{\otimes n}(u_n^{\otimes n}
v_{n}^{\otimes n}).
$$
\end{lemma}

\noindent{\it Proof.} We will use the Dyson expansion \cite{Ohya&Petz}
$$
\exp\{it(\log D+b)\}D^{-it}=\sum_{k=0}^\infty i^k
\int_0^tds_1\dots\int_0^{s_{k-1}}ds_k \sigma^D_{s_k}(b)\dots\sigma^D_{s_1}(b),
$$
where $\sigma^D_s(b)=D^{is}bD^{-is}$.  Let us denote $b_n=a_n-\varphi(a)$.
We get
\begin{eqnarray*}
&&
\varphi^{\otimes n}(u_n^{\otimes n}\, w_{n,t}^{\otimes n} \, v_{n}^{\otimes n})=\\
&&
\exp\{it\varphi(a)\} \varphi^{\otimes n}
\left(u_n^{\otimes n}\exp\{it(\log (\tilde{\rho}_n^{\otimes n})+S_n(b_n))\}
(\tilde{\rho}_n^{\otimes n})^{-it}v_n^{\otimes n}\right)=\\
&&\exp\{it\varphi(a)\} [\varphi^{\otimes n}
\left(u_{n}^{\otimes n}v_{n}^{\otimes n}\right)+ \\
&&  \varphi^{\otimes n}
(u_n^{\otimes n}
\sum_{k=1}^\infty i^k
\int_0^t ds_1\dots\int_0^{s_{k-1}} ds_k
S_n(\sigma^{\tilde{\rho}_n}_{s_k}(b_n))\dots
S_n(\sigma^{\tilde{\rho}_n}_{s_1}(b_n))v_n^{\otimes n})].
\end{eqnarray*}

The term in the last line can be rewritten as
\begin{eqnarray*}
&&\int_0^tds_1\sum_{k=0}^\infty\int_0^{s_1}dx_1\dots
\int_0^{x_{k-1}}dx_k \,i^{k}\times\\
&&\times \left\<S_n(\sigma^{\tilde{\rho}_n}_{x_1}(b_n))\dots
S_n(\sigma^{\tilde{\rho}_n}_{x_k}(b_n))(u_n^*)^{\otimes n}\Psi,\
iS_n(\sigma^{\tilde{\rho}_n}_{s_1}(b_n))v_n^{\otimes n}\Psi\right\>=\int_0^t d s_{1}
\times\\
&&\left\<(v_n^*\tilde{\rho_n}^{it})^{\otimes n}
\exp\{-is_1(\log
(\tilde{\rho}_n^{\otimes n})+S_n(b_n))\}(u_n^*)^{\otimes n}\Psi,\
iS_n(v_n^*\sigma^{\tilde{\rho}_n}_{s_1}(b_n)v_n)\Psi\right\>.
\end{eqnarray*}
The sequence
$v_n^*\sigma^{\tilde{\rho}_n}_{s}(b_n)v_n$ converges to
$\sigma^\varphi_s(a-\varphi(a))$ in norm and
$S_n(\sigma^\varphi_s(a-\varphi(a)))\Psi$ converges to 0,
 by the weak law of large numbers.
 Moreover, for all $n$ and $s$ we have
 $\|S_n(v_n^*\sigma^{\tilde{\rho}_n}_{s}(b_n)v_n)\|\le\|b_n\|$ and $\|b_n\|$
 is bounded. So the last term  goes to 0 as $n\to \infty$, by the dominated
 convergence theorem.

\qed

Let us now return to the family $\{\rho_{\theta} : \theta\in\Theta\}$, and consider the spectral decomposition $\rho_{\theta}:=\sum_j\lambda_{j,\theta}P_{j,\theta}$. By the differentiability of the map $\theta\mapsto \rho_{\theta}$ there exist
self-adjoint matrices $H_{j,\theta}\in \mathcal A$,
such that
\begin{equation}\label{def.H}
\frac {\partial}{\partial \theta_k} P_{j,\theta}=i[H_{k,\theta},P_{j,\theta}]
\qquad \theta\in\Theta,\ j=1,\dots, d,\ k=1\dots,m.
\end{equation}
We fix a point $\theta_0\in \Theta$ and make the notations $\rho=\rho_{\theta_0}$,
$P_j=P_{j,\theta_0}$,  $H_k=H_{k,\theta_0}$, and  $\tau_\theta=\sum_j\lambda_{j,\theta} P_j$.
For a smooth function  $f:\mathbb R\to \mathbb R$, we have
$$
\left. \frac {\partial}{\partial \theta_k} f(\rho_{\theta}) \right|_{\theta=\theta_{0}}=
\left.\frac{\partial}{\partial \theta_k}f(\tau_\theta) \right|_{\theta=\theta_{0}}+
 i[H_{k},f(\rho)],
\qquad \ k=1,\dots, m.
$$
The first term commutes with $\rho$ and the second term satisfies
$\Tr\, a[H_k,f(\rho)]=0$, whenever $[a,\rho]=0$. We may
suppose that $\varphi(H_k)=0$ for all $k$.


We will deal with expressions of the form
$\varphi^{\otimes n}(v_{n,1}^{\otimes n}\dots v_{n,k}^{\otimes n})$, where
\begin{equation}\label{eq.vnj}
v_{n,j} =\rho_{\theta_0+\frac {1}{\sqrt n}u^{j}}^{it_j}\rho^{-it_j},
\quad
\mathrm{or}
\quad
v_{n,j}=\rho^{it_j} \rho_{\theta_0+\frac 1{\sqrt n}u^j}^{-it_j}, \qquad
t_j\in\mathbb R, u^j\in I. 
\end{equation}
We will first show that the original family of states can be replaced by a
simpler one without changing the asymptotics.
\begin{lemma}\label{lemma.tilderho}
Let
$$
\tilde{\rho}_{a}=\exp\left(i \sum_ka_kH_k\right)
\tau_{\theta_0+a}\exp\left(-i \sum_ka_kH_k\right), \quad a\in I,
$$
and let
$\tilde {\mathcal E}_n=(\mathcal A_n, \tilde{\rho}_{n,u}:=
\tilde{\rho}_{u/\sqrt{n}}^{\otimes n}: u\in I)$.
Then $\lim_n\omega_{\iE_n}(g)=\lim_n\omega_{\tilde{\iE}_n}(g)$, for all
$g\in G$.
\end{lemma}

\noindent{\it Proof.}
Let us denote $\tilde v_{n,j}$ the expression obtained from $v_{n,j}$ by
replacing $\rho_{\theta_0+u^j/\sqrt n}$ by $\tilde\rho_{u^j/\sqrt{n}}$,
$j=1,\dots,k$. We have to show that
$$
\lim_n\varphi^{\otimes n}(v_{n,1}^{\otimes n}\dots v_{n,k}^{\otimes n})=
\lim_n \varphi^{\otimes n}(\tilde v_{n,1}^{\otimes n}\dots \tilde v_{n,k}^{\otimes n}).
$$
Let
$\rho_n=\rho_{n,0}=\tilde{\rho}_{n,0}$.
Then
$$
\rho_{n,u}^{it}\rho_n^{-it}=\exp\{it(\log \tilde{\rho}_{u/\sqrt n}+
\log {\rho}_{\theta_0+u/\sqrt n} -\log  \tilde{\rho}_{u/\sqrt n} )
\}^{\otimes n}
\tilde{\rho}_{n,u}^{-it}
\tilde{\rho}_{n,u}^{it}\rho_n^{-it}
$$
By considering the Taylor expansion of the functions $s\mapsto \log
\tilde{\rho}_{su/\sqrt n}$ and $s\mapsto
\log \rho_{\theta_0+su/\sqrt n}$, we get
$$
\log {\rho}_{\theta_0+u/\sqrt n} -\log  \tilde{\rho}_{u/\sqrt n}=
\frac{1}{2}\left(
\left. \frac {d^2}{ds^2}\log {\rho}_{\theta_0+\frac{su}{\sqrt{n}}} \right|_{s=s'_n} -
\left. \frac {d^2}{ds^2}\log  \tilde{\rho}_{\frac{su}{\sqrt{n}}} \right|_{s=s''_n}
\right)
$$
with $s'_n, s''_n\in [0,1]$ and it can be shown by some computation that
the last expression is equal to $\frac 1n a_n$, where $a_n$ converges in norm
to
$$
a= -\frac{1}{2} \rho^{-1} [ [\rho , H(u) ] , H(u) ],
\qquad
H(u)= \sum_{k} u_{k} H_{k},
$$
satisfying  $\varphi(a)=0$, where we have used the fact that the states
$\varphi_{\theta}$ are faithful and thus $\rho_{\theta}$ is invertible.
The statement can be now proved by a repeated  use of Lemma \ref{lemma:pom}.

\qed

We introduce the following notations:
\begin{eqnarray}
&&
 l(u):=\sum_k u_{k} l_{k} =\sum_{k} u_k\frac{\partial}{\partial \theta_k}
\log \tau_\theta|_{\theta=\theta_0},
 \label{def.f}\\
&&
h(u):= \sum_{k,l} u_k u_l \frac{\partial^2}
{\partial\theta_k\partial\theta_l}\log\tau_\theta|_{\theta=\theta_0},
\nonumber\\
&&
\ell (u) := \sum_k u_{k} \ell_{k},  \quad \ell_{k} \circ \rho = i[H_{k}, \rho],
\nonumber\\
&&
\mathcal{L} (u) := \sum_k u_{k} \mathcal{L}_{k},  \quad \mathcal{L}_{k} \circ \rho =
\frac{\partial}{\partial \theta_{k}} \rho_{\theta}|_{\theta=\theta_0}\label{def.ell}.
\end{eqnarray}
Note that $l_{k}$ are the logarithmic derivatives in $\theta_{0}$ of the
commutative family of states $\tau_{\theta}$. Similarly $ \ell_{k}$ is the symmetric logarithmic derivative of the unitary family obtained by rotating $\rho$ with the unitary. The sum $\mathcal{L}_{k}=l_{k}+\ell_{k}$ is the symmetric logarithmic derivative at $\theta_{0}$ of the original family $\rho_{\theta}$.
We notice further that $\varphi(H(u))=\varphi(l(u))=\varphi(\ell(u))=0$ and $-\varphi(h(u))=\varphi(l(u)^2)$
is the Fisher information of the family
$s\mapsto \tau_{\theta_0+su}$, at $s=0$.


We compute now the Connes cocycles for the family $\tilde{\rho}_{n,u}$:
\begin{eqnarray*}
&&
\tilde{\rho}_{n,u}^{it} \tilde{\rho}_n^{-it}=
\left[\exp\left(\frac i{\sqrt{n}}H(u)\right)
\tau_{\theta_0+\frac{1}{\sqrt n}u}^{it}\exp\left(-\frac i{\sqrt n}H(u)\right)\right]^{\otimes n}
(\rho^{-it})^{\otimes n}\\
&&=\exp\left(\frac i{\sqrt n}H(u)\right)^{\otimes n}
\left(\tau_{\theta_0+\frac{1}{\sqrt n}u}^{it}\rho^{-it}\right)^{\otimes n}
\left(\rho^{it}\exp\left(-\frac i{\sqrt n}H(u)\right)
\rho^{-it}\right)^{\otimes n}\\
&&= \exp\left(\frac i{\sqrt n}H(u)\right)^{\otimes n}
\exp\left(it(\log \tau_{\theta_0+u/\sqrt n} - \log\rho)\right)^{\otimes n}
\exp\left(-\frac i{\sqrt n}\sigma^\varphi_t(H(u))\right)^{\otimes n}
\end{eqnarray*}
Note that $\tau_{\theta_0}=\rho$ and all the elements
$\tau_\theta$ are mutually commuting.
Using again Taylor expansion up to the second order, we get
\begin{eqnarray*}
&&
\tilde{\rho}_{n,u}^{it}\tilde{\rho}_n^{-it}=\\
&&
=\exp\left(\frac i{\sqrt n}H(u)\right)^{\otimes n}
\exp\left(\frac {it}{\sqrt n}l(u) + \frac {it}{2n}b_n\right)^{\otimes n}
\exp\left(-\frac i{\sqrt n} \sigma^\varphi_t(H(u))\right)^{\otimes n}\\
&&=\exp\left(iF_n(H(u))\right)\exp\left(it(F_n(l(u))\right)\exp\left(\frac{it}2 S_n(b_n)\right)
\exp\left(-iF_n(\sigma^\varphi_t(H(u)))\right)
\end{eqnarray*}
where
$$
b_n=\sum_{k,l}u_ku_l\frac{\partial^2}{\partial\theta_k\partial \theta_l}
\log\tau_{\theta}|_{\theta=\theta_n},\qquad \|\theta_n-\theta_0\|\le
\frac 1{\sqrt n} \|u\|.
$$
By continuity of the second derivatives, $\{b_n\}$ converges to $h(u)$ in norm.
By the quantum Central Limit Theorem and
Lemma \ref{lemma:pom}, we can now conclude that the family of cocycles of the modified states $ \tilde{\rho}_{n,u}^{it}\tilde{\rho}_n^{-it}$ converges to
$$
V_{u,t}:=\exp\left(\frac{it}{2}\varphi(h(u))\right) W(H(u)) W(tl(u)) W(-\sigma^\varphi_t(H(u))),
$$
where $W(A)$ are the Weyl operators.
The convergence holds as usually in the weak sense: for any
$u_{1}, \dots u_{k}\in I$ and $t_{1}, \dots t_{k}\in\mathbb{R}$
$$
\lim_{n} \varphi^{\otimes n} (\tilde{v}_{n,1}^{\otimes n} \dots \tilde{v}_{n,k}^{\otimes n}) =
\phi( V_{1}\dots V_{k}),
 $$
where $V_{j}$ is shorthand notation for $V_{u_{j},t_{j}}$ or  $V_{u_{j},t_{j}}^{*}$, according to \eqref{eq.vnj}. In combination with Lemma \ref{lemma.tilderho} this  gives
 $$
 \lim_{n} \varphi^{\otimes n} (v_{n,1}^{\otimes n} \dots v_{n,k}^{\otimes n}) =
\phi( V_1\dots V_k).
$$
 It remains now to identify $V_{u,t}$ as Connes cocycles of the limit experiment,
$V_{u,t} = [D\phi^{u}, D\phi^{0}]_{t}$ where $\phi^{u}$ are states on the
algebra $CCR(M(\mathbb{C}^{d})^{sa},\sigma)$. Using the fact that $[B(l(u)) , B(A)] = 0$  for any $A\in M (\mathbb{C}^{d})^{sa}$ we can decompose $V_{u,t}$ into a product
\begin{eqnarray}\label{eq.cocycle.expression}
V_{u,t} &=&
 W\left(H(u)-\sigma_{t}(H(u)) )\right)
\exp
\left[\frac{1}{2} \varphi \left(
[H(u),\sigma_{t} (H(u))] \right)\right] \times \nonumber\\
&&
W\left( tl(u)\right) \exp
\left[\frac{it}{2} \varphi \left(  h(u) \right) \right].
\end{eqnarray}
where the first term is exactly the cocycle appearing in (\ref{eq.cocycle.unitary}) for the unitary family of states and the second term is the `classical cocycle' due to the change in the eigenvalues of the density matrix. We will show that indeed the product of
cocycles can be accounted for by a product of transformations such that
\begin{eqnarray*}
[D\phi^{u}, D\phi]_{t}
=
\left[D\left(\phi \circ R(u)\circ L(u)\right), D\left(\phi \circ R(u)\right)\right]_{t} \,
\left[D\left(\phi  \circ R(u)\right) , D\phi \right]_{t} .
\end{eqnarray*}
The {\it inner} automorphism $R(u)$ of $CCR(M(\mathbb{C}^{d})^{sa},\sigma)$ is the
`translation' with momentum $B(H)$
\begin{eqnarray}\label{eq.R}
R(u) : W(A) &\mapsto& W (H(u)) W(A)W (-H(u))\nonumber  \\
&=&
 W(A) \exp\{ i(A, \ell(u) )_{\rho}\},
\end{eqnarray}
just like in the unitary case (see eq. \eqref{def.ell}). The transformation $L(u)$ is an {\it outer} automorphism of $CCR(M(\mathbb{C}^{d})^{sa},\sigma)$, i.e. whose generator is a field which does not belong to the algebra as it corresponds to a non-selfadjoint operator
\begin{eqnarray}\label{eq.L}
L(u) :
 W(A) &\to & W ( -i l(u)/2) W(A)W (il(u)/2)
\nonumber\\
&=&
W(A) \exp \left\{ i ( A, l(u) )_{\rho}\right\}.
\end{eqnarray}
Using the factorization (\ref{eq.factorization.1}) of $CCR(M(\mathbb{C}^{d})^{sa},\sigma)$ and the definitions of $L(u)$ and $R(u)$ we get the following picture of the action the
product $L(u)\circ R(u)$ :
\begin{equation*}
L(u) \circ R(u) : W (B_{1}) \otimes W(B_{2}) \mapsto
 L(u) ( W(B_{1}) )  \otimes R(u) ( W(B_{2})) .
\end{equation*}
Moreover, from (\ref{eq.factorization.2}) we obtain that the state $\phi^{u}$ factorizes as well
$$
\phi^{u} = \phi_{1} \circ L(u) \otimes \phi_{2} \circ R(u) := \phi_{1}^{u} \otimes \phi_{2}^{u}.
$$
It is now easy to see that the cocycles for this family of states have the expression (\ref{eq.cocycle.expression}) and the states $\phi^{u}$ are given by \cite{Hayashi&Matsumoto}
\begin{eqnarray}\label{eq.phiu}
\phi^{u} \left(W (A) \right) &&
= \exp\left(- \frac{1}{2} (A,A)_{\rho} \right)
\exp
\left[
i \mathrm{Tr}\left(
A \sum_{i} u_{i}\frac{\partial \rho^{\theta}}{\partial \theta_{i}}
\right)
\right] \nonumber\\
&&
=  \exp\left(- \frac{1}{2} (A,A)_{\rho} + i  (A,  \mathcal{L}(u))_{\rho}\right) .
\end{eqnarray}
\begin{theorem}\label{th.qlan}
The sequence
$$
\mathcal{E}_{n} := \left( M(\mathbb{C}^{d})^{\otimes n},\varphi_{\theta_{0} + u/\sqrt{n}}^{\otimes n} : u\in I \right),
$$
of quantum statistical experiments converges weakly  as $n\to\infty$ to the limit experiment
$$
\mathcal{E}:= \left( CCR( M(\mathbb{C}^{d})^{sa} ,\sigma) , \phi^{u} : u \in I\right).
$$
The latter is a tensor product between a classical Gaussian shift experiment corresponding to the change in the eigenvalues of $\rho_{\theta}$, and a
non-commutative one corresponding to the rotation of the eigenbasis of $\rho_{\theta}$. On the algebraic level we have the isomorphism
$$
CCR\left( M(\mathcal{C}^{d})^{sa},\sigma \right)\cong CCR\left(\mathcal{H}_{\rho}\right) \otimes CCR\left(\mathcal{H}_{\rho}^{\perp}\right) ,
$$
as described in Section \ref{sec.qclt}.
With respect to this isomorphism the state $\phi^{u}$ given by \eqref{eq.phiu},
factorizes as
$$
\phi^{u} = \phi_{1}^{u} \otimes \phi_{2}^{u} = \phi_{1}\circ L(u) \otimes \phi_{2} \circ R(u),
$$
with automorphisms $R(u), L(u)$ defined in (\ref{eq.R}) and (\ref{eq.L}) respectively.
\end{theorem}


In the reminder of this section we will identify the minimal sufficient algebra $\mathcal{W}_{0}\subset CCR(M(\mathbb{C}^{d})^{sa},\sigma)$ of the experiment $\mathcal{E}$.
We know that the Connes cocycles generate the minimal sufficient algebra, and from the expression (\ref{eq.cocycle.expression}) we get that
$\mathcal{W}_{0} = CCR(K)$ where $K$ is the real linear space
$$
K:=   \mathrm{Lin}_{\mathbb{R}} \left\{H(u)-\sigma_{t}^{\varphi}(H(u)) +
t l(u) : t\in\mathbb{R} , u\in I\right\}.
$$
By taking derivatives with respect to $t$ and using the equations \eqref{def.H} and \eqref{def.f} we get that $K$ is the linear span of the orbits of the logarithmic derivatives
$\log\rho_{k}^{\prime}:=\partial \log \rho_{\theta}/\partial \theta_{k}|_{\theta=\theta_{0}}$ under the modular group $\sigma_{t}^{\varphi}$.
\begin{lemma}
The minimal sufficient algebra of the experiment $\mathcal{E}$ is given by
$\mathcal{W}_{0} = CCR(K, \sigma)$ with
$$
K = \mathrm{Lin}_{\mathbb{R}}
\left\{
l(u): u\in I \right\} \oplus
 \mathrm{Lin}_{\mathbb{R}}
\left\{
H(u) -\sigma_{t}^{\varphi} (H(u)):  u\in I , t\in\mathbb{R}
\right\}.
$$
In particular $\mathcal{L}_{k}\in K$ and $l_{k}\in K$.
\end{lemma}

\noindent{\it Proof.}
We have
 $$
 \mathcal{L}_{k} \circ \rho :=
 \left.\frac{\partial \rho_{\theta}}{\partial\theta_{k}}\right|_{\theta=\theta_{0}} =
 \sum_{i} \left. \frac{\partial\lambda_{i,\theta}}{ \partial\theta_{k}} \right|_{\theta=\theta_{0}} P_{i} +
\sum_{i} i \lambda_{i} [H_{k}, P_{i}],
 $$
which on the matrix elements becomes
\begin{equation}\label{eq.sld3}
\langle e_{i} ,\mathcal{L}_{k} e_{j} \rangle =
\delta_{ij} \left. \frac{\partial\log \lambda^{\theta}_{i}}{\partial\theta_{k}}\right\vert_{\theta=\theta_{0}}
+
2i \frac{1- e^{\log(\lambda_{j}/\lambda_{i})}}{1+e^{\log(\lambda_{j}/\lambda_{i})}}
\langle e_{i} ,H_{k} e_{j} \rangle.
\end{equation}
The logarithmic derivative $\log\rho_{k}^{\prime}$ is in $K$ and has matrix elements
$$
\langle e_{i} ,\log\rho_{k}^{\prime} e_{j} \rangle =
\delta_{ij} \left. \frac{\partial\log \lambda^{\theta}_{i}}{\partial\theta_{k}}\right|_{\theta=\theta_{0}} +i \log(\lambda_{j}/\lambda_{i})\langle e_{i}, H_{k} e_{j} \rangle,
$$
and by derivating $\sigma_{t}\left(\log\rho_{k}^{\prime}\right)$ we get that the multiple commutators
$$
C_{r} =  [\dots[ \log\rho_{k}^{\prime}, \log\rho], \dots, \log \rho],
$$
 are also in $K$ and have the expression
$$
\langle e_{i} , C_{r} e_{j} \rangle =
i \log(\lambda_{j}/\lambda_{i})^{r+1}\langle e_{i}, H_{k} e_{j} \rangle .
$$
for $r\geq 1$. By comparing with (\ref{eq.sld3}) with the last two equations we conclude that $\mathcal{L}_{k}\in K$ for all $k=1,\dots ,m$, and additionally
that $l(u)\in K$ for all $u\in I$. Indeed, there exist a finite number of real
coefficients $\{a_{r}, r=0 \dots   \} $ such that
$$
\sum_{r=0}^{d(d-1)} a_{r} \log(\lambda_{j}/\lambda_{i})^{r} =0 , \forall  1\leq i\neq j\leq d,
$$
and $a_{0} \neq 0$. With such coefficients we have
$$
a_{0} \log \rho_{k}^{\prime} +
\sum_{r=1}^{d(d-1)} a_{r} C_{r} = a_{0} l_{k} \in K.
$$
In conclusion $K$ is the linear span of the vectors
$l_{k}\in\mathcal{H}_{\rho}$ and the vectors
$ H_{k}-\sigma_{t}(H_{k}) \in  \mathcal{H}_{\rho}^{\perp}$ as desired.

\qed

Another interesting feature of the minimal sufficient algebra $W_{0}$ is that apart from the standard symmetric logarithmic derivatives $\mathcal{L}_{k}$, it contains
a broad set of quantum versions of the logarithmic derivative which were investigated
in \cite{Petz2} and are defined as follows
$$
\mathcal{L}^F_{k}=
J^F\left(  \rho_{k}^{\prime} \right),
$$
where
$\rho_{k}^{\prime}=\partial \rho_{\theta}/\partial\theta_{k}|_{\theta=\theta_0}$
and $J^F$ is an operator on matrices defined as
$$
J^F =
[F(L R^{-1})]^{-1}R^{-1}.
$$
Here $L$ and $R$ are the left and respective right multiplication by $\rho$, and $F:\mathbb R^+\to \mathbb R$ is an operator-monotone function satisfying
$F(t)=tF(t^{-1})$ for $t>0$ and $F(1)=1$.
This function is required to satisfy the physical admissibility condition
that the associated quantum Fisher information
$I_{kp}:=\Tr  \rho_{k}^{\prime}\mathcal L^F_{p}$ is monotone
under coarse-grainings. Two well-known examples of a quantum score are the
symmetric logarithmic derivative $\mathcal{L}_k$, for which $F(t)=(1+t)/2$,
and the Bogoljubov-Kubo-Mori logarithmic derivative  $\mathcal
{L}^{BMK}_k:=\log \rho_{k}^{\prime}$ for which $F(t)=\frac{t-1}{\log(t)}$, and as we
have seen they both belong to the subspace $K$.
\begin{lemma}
For any admissible function $F$ the logarithmic derivative
$\mathcal{L}^{F}_{k}$ belongs to $K$.
\end{lemma}

\noindent{Proof.}
First, we see that
$$
\langle e_{i} ,\mathcal{L}^F_{k} e_{j} \rangle =
\delta_{ij} \left. \frac{\partial\log \lambda^{\theta}_{i}}{\partial\theta_{k}}\right|_{\theta=\theta_{0}}
+ i\frac{(1-\lambda_i/\lambda_j)}{F(\lambda_i/\lambda_j)}
\langle e_i, H_{k} e_j\rangle.
$$
Furthermore, for each $F$ we have  the integral representation  \cite{Lesniewski&Ruskai}
$$
\frac{1-t}{F(t)}= \int_{[0,\infty]}\frac {1-t}{s+t} (1+s)\mu(ds),
$$
where $\mu(s)$ is a positive finite  measure on $[0,\infty]$.
Therefore,
$$
\langle e_{i} ,\mathcal{L}^F_{k} e_{j} \rangle =
\delta_{ij} \left. \frac{\partial\log \lambda^{\theta}_{i}}{\partial\theta_{k}}\right|_{\theta=\theta_{0}}
+ i\int_{[0,\infty]}
\frac{(1-\lambda_i/\lambda_j)}{(s+\lambda_i/\lambda_j)}
\langle e_i, H_{k} e_j\rangle(1+s)\mu(ds),
$$
and $\mathcal L^F_{k}\in K$ is proved similarly as for $\mathcal
L_{k}$.

\qed

\section{Application to qubit states}
In this section we apply the local asymptotic normality results to the simplest situation of
a family of qubit states. In Theorem 1.1 of \cite{Guta&Kahn} it is shown that in this case local asymptotic normality holds in the strong sense of Definition
\ref{def.strong.convergence}.

An arbitrary density matrix in $M(\mathbb{C}^{2})$ can be written as
$$
\rho = \frac{\mathbf{1} + \overrightarrow r \overrightarrow \sigma }{2}  ,
$$
where $\overrightarrow r = (r_{x}, r_{y}, r_{z})\in \mathbb{R}^{3}$ is a vector
satisfying $|\overrightarrow r|\le 1$, and $\overrightarrow \sigma=(\sigma_{x}, \sigma_{y}, \sigma_{z})$ are the Pauli matrices. Due to the rotation symmetry, we
may choose  $\rho_{0} = \frac{\mathbf{1} + r\sigma_{z}}{2}$ corresponding to
$ \overrightarrow{r_{0}} = (0,0,r)$ for some fixed $r\in (0,1)$. All the states in a neighborhood of $\rho_{0}$ can be obtained by a combination of a translation in the radial direction, and a rotation around an axis in the $x$-$y$ plane.
Thus we can use the local coordinates
$\overrightarrow u= (r_{x},r_{y}, a)$ around $\overrightarrow{r_{0}}$ such that
$$
\rho_{\overrightarrow u} =
\frac{\mathbf{1} +( \overrightarrow{r_{0}} +  \overrightarrow{u}) \overrightarrow \sigma}{2}.
$$
Notice that only the coordinate $a$ contributes to the classical part of the experiment
calculate and the functions $l$ and $h$ defined in Section \ref{sec.lan.general} are
\begin{eqnarray*}
l_{a} &=& \left. \frac{\partial \log\rho_{\overrightarrow{u}}}{\partial a}\right\vert_{\bf u=0} =
 \frac{1}{1+r} P_{+} - \frac{1}{1-r}P_{-},
\\
h_{aa} &=& \left. \frac{\partial^{2} \log\rho_{\overrightarrow{u}}}{\partial a^{2}}\right\vert_{\bf u=0}=
-\left(\frac{1}{(1+r)^{2}} P_{+} + \frac{1}{(1-r)^{2}}P_{-}\right),
\end{eqnarray*}
where $P_{\pm}$ are the eigenprojectors of $\rho_{0}$, and the components corresponding to other derivatives are equal to $0$. With the notations defined in Section \ref{sec.qclt}, we construct the real Hilbert space $L^{2}_{\mathbb{R}} (\rho_{0})$ with inner product
$$
(A, B)_{\rho_0} = \mathrm{Tr} \left(\rho_0 A\circ B \right), \qquad A, B\in M(\mathbb{C}^{2})^{sa},
$$
with respect to which we have the orthogonal decomposition
\begin{equation}\label{eq.orth.split}
L^{2}_{\mathbb{R}} (\rho_0) =
\mathcal{H}_{\rho_0}\oplus \mathcal{H}_{\rho_0}^{\perp}=
\mathrm{Lin} \{ \mathbf{1}, \sigma_{z}\} \oplus \mathrm{Lin}\{ \sigma_{x} , \sigma_{y}\}.
\end{equation}
Next, we use the symplectic form
$
\sigma(A,B) = \frac{i}{2}\mathrm{Tr}\left( \rho_0[A, B] \right) ,
$
to construct the algebra $CCR(M(\mathbb{C}^{2})^{sa},\sigma)$.
We obtain that $B(\sigma_{z})$ and $B(\mathbf{1})$ commute with all the other fields
and $B(\sigma_{y})$, $B(\sigma_{x})$ satisfy the canonical commutation relations
$$
[B(\sigma_{y}), B(\sigma_{x})] = 2i r \mathbf{1}.
$$
By rescaling we get the usual quantum oscillator relations
$[{\bf Q}, {\bf P}] = i\mathbf{1}$ with ${\bf Q} = B(\sigma_{y})/\sqrt{2r}$ and
${\bf P} = B(\sigma_{x})/\sqrt{2r}$.
Thus
$$
CCR( M(\mathbb{C}^{2})^{sa},\sigma ) \cong
CCR(\mathrm{Lin} \{ \mathbf{1}, \sigma_{z}\} )\otimes
{\rm Alg}( {\bf Q} , {\bf P}).
$$
where the left side of the tensor product is itself a commutative algebra which is naturally isomorphic to $L^{\infty} (\mathbb{R}^{2})$, and the right
side is the algebra of a quantum harmonic oscillator with variables ${\bf Q}$ and
${\bf P}$.
On this algebra we have a state $\phi^0$ given by
$
\phi^0 (W(A)) = \exp\left(-\frac{1}{2} \mathrm{Tr} (\rho_{\bf 0 } A^{2})\right),
$
which due to (\ref{eq.orth.split}) splits into a tensor product
$
\phi^0 = \phi^0_{1} \otimes \phi^0_{2}.
$
In Section \ref{sec.lan.general} we have shown that the minimal sufficient algebra of the limit experiment is generated by the fields corresponding to a real linear subspace $K\subset M(\mathbb{C}^{2})^{sa}$ which in this case is
$
K = \mathbb{R} l_{a} \oplus \mathrm{Lin} \{ \sigma_{x}, \sigma_{y}\} .
 $
Then the minimal sufficient algebra is of the form
$$
CCR(K,\sigma) \cong L^{\infty}(\mathbb{R})\otimes
{\rm Alg} ({\bf Q}, {\bf P})
$$
and the family of states defining the limit experiment is
$$
\phi^u = N(I_c a , I_{c}) \otimes \phi^{r_{x},r_{y}}_{2}.
$$
Let us explain the meaning of the right side:
$$
I_{c} = \mathrm{Tr}(\rho_0l_{a}^{2}) =
-  \mathrm{Tr}(\rho_0  h_{aa} )= \frac{1}{1-r^{2}},
$$
is the Fisher information corresponding to the parameter $a$.
The state $\phi^{r_{x},r_{y}}_{2}$ of the quantum oscillator can be described through its Wigner function \cite{Leonhardt}
$$
W^{r_{x}r_{y}} (q,p)= \exp\left[ - r \left( (q- q_{x} )^{2}+
                                                             (p-p_{y})^{2} \right)\right],
$$
which corresponds to a displaced thermal equilibrium state with center
$(q_{x},q_{y})= (r_{x}/\sqrt{2r},r_{y}/\sqrt{2r})$.

\section{Concluding remarks}

In this paper we have made a further step in the development of a theory of
quantum statistical experiments started by Petz.
We believe that the notions which we have introduced are the proper analogues of the classical concepts: weak and strong convergence of experiments, canonical state of an experiment, local asymptotic normality. However the theory is far from complete and the following is a short list of open problems and topics for future work.

\noindent
1. Extend the theory of statistical experiments to the case of non-faithful states. We expect that the extended space of experiments will be compact under the weak
topology.  

2. One of the crucial aspects of the theory is the relation between strong and weak convergence of experiments for finite parameter sets. 
In Theorem \ref{th.weak&strong} we have touched upon this by showing that the two notions are equivalent when the experiments are uniformly dominated and the corresponding algebras are of type I with discrete center. We believe that the same result holds for a much larger class of experiments, where one would have to 
consider non-trivial channels in order to achieve the convergence in Le Cam sense. One possibility, perhaps too ambitious, would be to construct a quantum version of the Skorohod almost sure representation Theorem \cite{Pollard}. Another strategy could be to approximate the quantum experiments by finite dimensional ones, similarly 
to the treatment of nuclear C$^{*}$-algebras \cite{Takesaki3}. 
%

3. The work on the previous issue might be simplified by finding alternative characterizations of weak convergence in terms of quantum Radon-Nikodym derivatives.

4. Derive local asymptotic normality under weaker smoothness conditions for the family of states, similar to the differentiability in quadratic mean from the classical set-up \cite{vanderVaart}. Going beyond the finite parameter, i.i.d. case -- which
classically is rather standard -- remains a challenge for the quantum theory.

5. Develop a quantum statistical decision theory for quantum experiments. This will connect the abstract framework to concrete statistical problems such as estimation and testing.


\noindent
{\it Acknowledgments.} We thank Richard Gill, Aad van der Vaart, Denes Petz and
Jonas Kahn for fruitful discussions. M\u{a}d\u{a}lin Gu\c{t}\u{a} acknowledges the financial support received from  the Netherlands Organisation for Scientific Research (NWO).  Anna Jen\v cov\'a was supported by the
Center of Excellence SAS Physics of Information I/2/2005,  Science and
Technology Assistance Agency under the contract  No. APVT-51-032002
and the EU
Research Training Network Quantum Probability with Applications to
Physics, Information Theory and Biology.


\begin{thebibliography}{10}

\bibitem{Mabuchi}
M.~A. {Armen}, J.~K. {Au}, J.~K. {Stockton}, A.~C. {Doherty}, and H.~{Mabuchi}.
\newblock {Adaptive Homodyne Measurement of Optical Phase}.
\newblock {\em Phys. Rev. Lett.}, 89:133602, 2002.

\bibitem{Gill&Guta&Artiles}
{Artiles, L}, {Gill, R.}, and {Gu{\c t}{\u a}, M.}
\newblock An invitation to quantum tomography.
\newblock {\em J. Royal Statist. Soc. B (Methodological)}, 67:109--134, 2005.

\bibitem{Bagan&Baig&Tapia}
E.~Bagan, M.~Baig, and R.~Munoz-Tapia.
\newblock Optimal scheme for estimating a pure qubit state via local
  measurements.
\newblock {\em Phys. Rev. Lett.}, 89:277904, 2002.

\bibitem{Bagan&Gill}
E.~Bagan, {Ballester, M. A.}, {Gill, R. D.}, {Monras, A.}, and {Mun\~ oz-Tapia,
  R.}
\newblock Optimal full estimation of qubit mixed states.
\newblock {\em Phys. Rev. A}, 73:032301, 2006.

\bibitem{Barndorff-Nielsen&Gill&Jupp}
O.~E. Barndorff-Nielsen, {Gill, R.}, and {Jupp, P.~E.}
\newblock On quantum statistical inference (with discussion).
\newblock {\em J. R. Statist. Soc. B}, 65:775--816, 2003.

\bibitem{Belavkin}
V.~P. Belavkin.
\newblock Generalized heisenberg uncertainty relations, and efficient
  measurements in quantum systems.
\newblock {\em Theor. Math. Phys.}, 26:213--222, 1976.

\bibitem{Butucea&Guta&Artiles}
C.~Butucea, M.~Gu\c{t}\u{a}, and L.~Artiles.
\newblock {Minimax and adaptive estimation of the Wigner function in quantum
  homodyne tomography with noisy data}.
\newblock arxiv.org/abs/math/0504058, to appear in Annals of Statistics {\bf
  35}, 2007.

\bibitem{Cirac}
J.~I. Cirac, A.~K. Ekert, and C.~Macchiavello.
\newblock Optimal purification of single qubits.
\newblock {\em Phys. Rev. Lett.}, 82:4344, 1999.

\bibitem{D'Ariano.2}
G.~M. D'Ariano, {Leonhardt, U.}, and {Paul, H.}
\newblock Homodyne detection of the density matrix of the radiation field.
\newblock {\em Phys. Rev. A}, 52:R1801--R1804, 1995.

\bibitem{DellÕAntonio}
G.~F. DellÕAntonio.
\newblock On the limits of sequences of normal states.
\newblock {\em Comm. Pure Appl. Math.}, 20:413Ð429, 1967.

\bibitem{Gill}
R.~D. Gill.
\newblock Asymptotic information bounds in quantum statistics.
\newblock quant-ph/0512443, to appear in Annals of Statistics.

\bibitem{Gill&Massar}
R.~D. Gill and S.~Massar.
\newblock State estimation for large ensembles.
\newblock {\em Phys. Rev. A}, 61:042312, 2000.

\bibitem{Guta&Janssens&Kahn}
M.~Gu\c{t}\u{a}, B.~Janssens, and J.~Kahn.
\newblock Optimal estimation of qubit states with continuous time measurements.
\newblock www.arxiv.org/quant-ph/0608074, 2006.

\bibitem{Guta&Kahn}
M.~Gu\c{t}\u{a} and J.~Kahn.
\newblock Local asymptotic normality for qubit states.
\newblock {\em Phys. Rev. A}, 73:052108, 2006.

\bibitem{Guta&Matsumoto}
M.~Gu\c{t}\u{a} and K.~Matsumoto.
\newblock Optimal cloning of mixed gaussian states.
\newblock {\em Phys. Rev. A}, 74:032305, 2006.

\bibitem{Hannemann&Wunderlich}
T.~Hannemann, D.~Reiss, C.~Balzer, W.~Neuhauser, P.~E. Toschek, and
  C.~Wunderlich.
\newblock {Self-learning estimation of quantum states}.
\newblock {\em Phys. Rev. A}, 65:050303--+, 2002.

\bibitem{Hayashi.conference}
M.~Hayashi.
\newblock presentations at maphysto and quantop workshop on quantum
  measurements and quantum stochastics, aarhus, 2003, and special week on
  quantum statistics, isaac newton institute for mathematical sciences,
  cambridge, 2004.

\bibitem{Hayashi.japanese}
M.~Hayashi.
\newblock Quantum estimation and the quantum central limit theorem.
\newblock {\em Bulletin of the Mathematical Society of Japan}, 55:368--391,
  2003.
\newblock ( in Japanese; Translated into English in quant-ph/0608198).

\bibitem{Hayashi&Matsumoto2}
M.~Hayashi and K.~Matsumoto.
\newblock Statistical model with measurement degree of freedom and quantum
  physics.
\newblock In Masahito Hayashi, editor, {\em Asymptotic theory of quantum
  statistical inference: selected papers}, pages 162--170. World Scientific,
  2005.
\newblock ({E}nglish translation of a paper in Japanese published in
  Surikaiseki Kenkyusho Kokyuroku, vol. 35, pp. 7689-7727, 2002.).

\bibitem{Hayashi&Matsumoto}
M.~Hayashi and {Matsumoto, K.}
\newblock Asymptotic performance of optimal state estimation in quantum two
  level system.
\newblock quant-ph/0411073.

\bibitem{Hayashi.editor}
Masahito Hayashi, editor.
\newblock {\em Asymptotic theory of quantum statistical inference: selected
  papers}.
\newblock World Scientific, 2005.

\bibitem{Hayashi.book}
Masahito Hayashi.
\newblock {\em Quantum Information}.
\newblock Springer-Verlag, Berlin Heidelberg, 2006.

\bibitem{Helstrom69}
C.~W. Helstrom.
\newblock {Quantum detection and estimation theory}.
\newblock {\em Journal of Statistical Physics}, 1:231--252, 1969.

\bibitem{Holevo}
A.~S. Holevo.
\newblock {\em Probabilistic and Statistical Aspects of Quantum Theory}.
\newblock North-Holland, 1982.

\bibitem{Kadison&Ringrose}
R.~V. Kadison and J.~R. Ringrose.
\newblock {\em Fundamentals of the Theory of Operator Algebras I}.
\newblock American Mathematical Society, 1997.

\bibitem{Keyl&Werner}
M.~Keyl and R.~F. Werner.
\newblock Estimating the spectrum of a density operator.
\newblock {\em Phys. Rev. A}, 64:052311, 2001.

\bibitem{Koashi&Imoto}
M.~{Koashi} and N.~{Imoto}.
\newblock {Operations that do not disturb partially known quantum states}.
\newblock {\em Phys. Rev. A}, 66:022318, 2002.

\bibitem{LeCam}
L.~Le~Cam.
\newblock {\em Asymptotic Methods in Statistical Decision Theory}.
\newblock Springer Verlag, New York, 1986.

\bibitem{Leonhardt}
U.~Leonhardt.
\newblock {\em Measuring the Quantum State of Light}.
\newblock Cambridge University Press, 1997.

\bibitem{Leonhardt.Munroe}
U.~Leonhardt, M.~Munroe, T.~Kiss, Th. Richter, and M.~G. Raymer.
\newblock Sampling of photon statistics and density matrix using homodyne
  detection.
\newblock {\em Optics Communications}, 127:144--160, 1996.

\bibitem{Lesniewski&Ruskai}
A.~Lesniewski and {Ruskai, M.B.}
\newblock Monotone riemannian metrics and relative entropy on noncommutative
  probability spaces.
\newblock {\em J. Math. Phys.}, 40:5702--5724, 1999.

\bibitem{Massar&Popescu}
S.~Massar and S~Popescu.
\newblock Optimal extraction of information from finite quantum ensembles.
\newblock {\em Phys. Rev. Lett.}, 74:1259--1263, 1995.

\bibitem{Petz&Mosonyi}
M.~{Mosonyi} and D.~{Petz}.
\newblock {Structure of sufficient quantum coarse-grainings}.
\newblock {\em Lett. Math. Phys.}, 68:19--30, 2004.

\bibitem{Nagaoka&Ogawa}
T.~Ogawa and H.~Nagaoka.
\newblock On the statistical equivalence for sets of quantum states.
\newblock {UEC}-IS-2000-5, IS Technical Reports, Univ. of Electro-Comm., 2000.

\bibitem{Ohya&Petz}
M.~Ohya and {Petz, D.}
\newblock {\em Quantum Entropy and its Use}.
\newblock Springer Verlag, Berlin-Heidelberg, 2004.

\bibitem{Paris.editor}
M.~G.~A. Paris and J.~{\v R}eh{\'a}{\v c}ek, editors.
\newblock {\em {Quantum State Estimation}}, 2004.

\bibitem{Petz86}
D.~Petz.
\newblock Sufficient subalgebras and the relative entropy of states of a von
  neumann algebra.
\newblock {\em Commun. Math. Phys.}, 105:123--131, 1986.

\bibitem{Petz}
D.~Petz.
\newblock {\em An Invitation to the Algebra of Canonical Commutation
  Relations}.
\newblock Leuven University Press, 1990.

\bibitem{Petz2}
D.~{Petz}.
\newblock {Covariance and Fisher information in quantum mechanics}.
\newblock {\em Journal of Physics A Mathematical General}, 35:929--939, 2002.

\bibitem{Petz&Jencova}
D.~Petz and {Jencova, A.}
\newblock Sufficiency in quantum statistical inference.
\newblock {\em Commun. Math. Phys.}, 263:259 -- 276, 2006.

\bibitem{Pollard}
D.~Pollard.
\newblock {\em Convergence of Stochastic Processes}.
\newblock Springer-Verlag, 1984.

\bibitem{Breitenbach&Schiller&Mlynek}
S.~Schiller, G.~Breitenbach, S.~F. Pereira, T.~M\"{u}ller, and J.~Mlynek.
\newblock Quantum statistics of the squeezed vacuum by measurement of the
  density matrix in the number state representation.
\newblock {\em Phys. Rev. Lett.}, 77:2933--2936, 1996.

\bibitem{Smith&Silberfarb&Deutsch&Jessen}
G.~A. Smith, A.~Silberfarb, I.~H. Deutsch, and P.~S. Jessen.
\newblock {Efficient Quantum-State Estimation by Continuous Weak Measurement
  and Dynamical Control}.
\newblock {\em Phys. Rev. Lett.}, 97:180403--+, 2006.

\bibitem{Smithey}
D.~T. Smithey, {Beck, M.}, {Raymer, M.~G.}, and {Faridani, A.}
\newblock Measurement of the {Wigner} distribution and the density matrix of a
  light mode using optical homodyne tomography: Application to squeezed states
  and the vacuum.
\newblock {\em Phys. Rev. Lett.}, 70:1244--1247, 1993.

\bibitem{Strasser}
H.~Strasser.
\newblock {\em Mathematical Theory of Statistics}.
\newblock De Gruyter, Berlin, New York, 1985.

\bibitem{Ta.}
M.~Takesaki.
\newblock {\em Theory of Operator Algebras I}.
\newblock Springer Verlag, New York, 1979.

\bibitem{Takesaki2}
M.~Takesaki.
\newblock {\em Theory of Operator Algebras II}.
\newblock Springer Verlag, Berlin, 2003.

\bibitem{Takesaki3}
M.~Takesaki.
\newblock {\em Theory of Operator Algebras III}.
\newblock Springer Verlag, Berlin, 2003.

\bibitem{Torgersen}
E.~Torgersen.
\newblock {\em Comparison of Statistical Experiments}.
\newblock Cambridge University Press, 1991.

\bibitem{vanderVaart}
A.W. van~der Vaart.
\newblock {\em Asymptotic Statistics}.
\newblock Cambridge University Press, 1998.

\bibitem{vanderVaart&Wellner}
A.W. van~der Vaart and {Wellner, J.A.}
\newblock {\em Weak Convergence and Empirical Processes}.
\newblock Springer, New York, 1996.

\bibitem{Vidal}
G.~Vidal, J.~I. Latorre, P.~Pascual, and R.~Tarrach.
\newblock Optimal minimal measurements of mixed states.
\newblock {\em Phys. Rev. A}, 60:126, 1999.

\bibitem{Vogel&Risken}
K.~Vogel and {Risken, H.}
\newblock Determination of quasiprobability distributions in terms of
  probability distributions for the rotated quadrature phase.
\newblock {\em Phys. Rev. A}, 40:2847--2849, 1989.

\bibitem{Wald}
A.~Wald.
\newblock {\em Statistical Decision Functions}.
\newblock John Wiley $\&$ Sons, New York, 1950.

\bibitem{Yuen&Lax&Kennedy}
H.~Yuen, R.~Kennedy, and M.~Lax.
\newblock {Optimum testing of multiple hypotheses in quantum detection theory}.
\newblock {\em IEEE Trans. Inform. Theory}, 21:125-- 134, 1975.

\bibitem{Yuen&Lax}
H.~P. Yuen and {Lax, M.}
\newblock Multiple-parameter quantum estimation and measurement of
  non-selfadjoint observables.
\newblock {\em IEEE Trans. Inform. Theory}, 19:740, 1973.

\bibitem{Zavatta}
A.~Zavatta, S.~Viciani, and M.~Bellini.
\newblock Quantum to classical transition with single-photon-added coherent
  states of light.
\newblock {\em Science}, 306:660--662, 2004.

\end{thebibliography}

\end{document}